# Physics-Informed Hemodynamic Modeling for Data-Free Prediction and Sparse-Data Assimilation

Xi Chen[1], Jianchuan Yang[2], Hongde Li[5], Guangxin He[3], Qiuyu Ye[4], Qiang Luo[5]†, Mao Chen[5]†, Wenqi Hu[1]†

1. Department of Mechanical and Aerospace Engineering, The Hong Kong University of Science and Technology, Clear Water Bay, Kowloon, Hong Kong, China.
2. Department of Civil and Environmental Engineering, The Hong Kong University of Science and Technology, Clear Water Bay, Kowloon, Hong Kong, China.
3. Department of Computer Science and Engineering, The Hong Kong University of Science and Technology, Clear Water Bay, Kowloon, Hong Kong, China.
4. Department of Electrical and Electronic Engineering, The Hong Kong Polytechnic University, Kowloon Hung Hom, Hong Kong, China.
5. Department of Cardiology, Laboratory of Cardiac Structure and Function at Institute of Cardiovascular Diseases, and Cardiac Structure and Function Research Key Laboratory of Sichuan Province, West China Hospital, Sichuan University, No.37 Guoxue Street, Chengdu 610041, P.R. China.

†Corresponding author

## Abstract:

Clinical decision-making for coronary intervention relies mainly on angiography and fractional flow reserve (FFR). However, angiography is two-dimensional and lacks depth information for 3D lesion characterization, while FFR provides only a single functional index, offering limited hemodynamic insight. Among existing methods, numerical analysis is computationally expensive, whereas learning-based approaches require extensive supervision and often lack physical consistency. To address these limitations, we propose physics-informed hemodynamic modeling, an integrated deep learning framework for 3D coronary blood flow analysis from dual-view angiography. First, an attention-enhanced CNN reconstructs coronary geometry from angiography. The resulting point clouds are then mapped to a reference domain and Fourier-encoded for joint representation. A decoupled network separately predicts velocity and pressure fields, with embedded physical priors enabling efficient transfer across physiological conditions. Across 32 clinical patients evaluated under four flow conditions, the trans-stenotic pressure-drop mean absolute percentage error was 2.02 ± 2.98%, while the velocity and pressure relative-L2 errors were 0.054 ± 0.023 and 0.023 ± 0.016, respectively. Validation against hospital-measured FFR further achieved 93.8% diagnostic accuracy (30/32; exact 95% CI, 79.2%–99.2%). The framework also supports illustrative revascularization comparisons and sparse-data assimilation, with the full angiography-to-hemodynamics pipeline completed within 20 minutes per patient.

# 1. Introduction

Coronary artery disease (CAD) remains a leading cause of mortality worldwide, driven by the accumulation of atherosclerotic plaques that induce localized coronary stenosis (Figure 1a) [1-3]. For clinical coronary assessment, X-ray coronary angiography is the most widely used imaging modality, providing real-time high-resolution projections for visualizing and quantifying luminal narrowing [4, 5]. However, its projection-based nature collapses 3D vascular anatomy onto planar images (Figure 1b), leading to the loss of depth information [6, 7]. Accurate interpretation therefore often requires repeated acquisitions from multiple viewing angles and subjective clinical inference, rendering assessments vulnerable to projection overlap, viewing-angle dependency, and local occlusion [8-13]. Moreover, increasing clinical evidence indicates that anatomical stenosis severity alone often fails to reflect functional significance, as similar degrees of narrowing can produce markedly different hemodynamic and perfusion outcomes [14]. Consequently, fractional flow reserve (FFR), a flow-based functional index defined by the pressure ratio across a coronary stenosis under hyperemia, as shown in Figure 1a, has been adopted as the clinical gold standard for assessing the physiological significance of coronary stenosis [14-18]. However, its invasiveness, procedural complexity, and associated cost still limit its widespread use, particularly in routine clinical practice and large-scale screening scenarios [19, 20]. Established angiography-derived approaches, including $FFR_{QCA}$ [21], QFR [22], vFFR [23, 24], caFFR [25], and μQFR [26], enable lesion-level FFR assessment and have undergone diagnostic validation. CT-FFR [18, 20] and stress perfusion CMR [27] offer complementary non-invasive anatomical–functional and perfusion assessments, respectively. However, methods for efficiently resolving patient-specific 3D velocity, pressure, and wall shear stress fields remain limited.

To overcome these limitations, extensive efforts have focused on 3D reconstruction methods for coronary anatomy and functional assessment. Early projection-based approaches relied on precise imaging system calibration and were highly sensitive to imaging parameters and viewing-angle accuracy, limiting their applicability in clinical practice where calibration errors can reach approximately 10% [28, 29]. Recent data-driven approaches learn the mapping between 2D angiographic views and 3D vascular geometry, making them more robust to calibration noise [30, 31]. Nevertheless, the resulting geometries still require conventional numerical solvers for subsequent hemodynamic analysis [32, 33]. The high computational cost and complex preprocessing steps such as mesh generation hinder their integration into routine clinical settings. Moreover, its deterministic pipeline lacks flexibility to incorporate clinical observations and typically cannot perform rapid cross-physiological evaluations, as each scenario requires re-computation from scratch. This absence of transferable learning capability constrains scalability and limits real-time clinical decision support (Figure 1c).

In recent years, physics-informed neural networks (PINNs) have emerged as a powerful paradigm for modeling complex physical systems [34-37]. By embedding governing equations and observational data directly into the training objective, PINNs have been successfully applied to various problems in hemodynamics, including modeling of cardiac valves [38, 39], coronary and other cardiovascular arteries [40, 41], and abdominal aortic aneurysms [42-44]. Yet in clinical practice, measurements are often sparse, and patient-specific high-quality flow annotations remain limited. This constraint necessitates PINNs that can infer physiologically consistent coronary flows from minimal or even unlabeled data, highlighting the importance of unsupervised physics-informed learning.

However, current PINN formulations face substantial challenges in complex hemodynamic modeling. Competition among multiple loss terms often leads to a complex, non-convex, and ill-conditioned optimization landscape (schematically illustrated in Figure 1c), resulting in slow convergence and unstable predictive accuracy [45-47]. In unsupervised settings, PINNs further exhibit spectral bias and struggle to capture high-frequency physical features, such as sharp pressure gradients and localized flow acceleration [48, 49]. When applied to coronary arteries with complex geometries, the strong coupling between intricate vascular structures and nonlinear flow dynamics produces stiff physical residuals [50, 51]. This stiffness impedes the effective propagation of physical constraints during training, substantially degrading computational efficiency and compromising solution reliability.

Moreover, coronary blood flow exhibits pronounced velocity variations across the cardiac cycle and under different physiological states, posing additional challenges for conventional PINNs. Under resting conditions, the mean velocity in the right coronary artery ranges from 0.144–0.209 m/s across diastole and systole [52, 53], while adenosine-induced maximal hyperemia during FFR assessment or exercise can elevate inlet velocity by approximately 2–5 fold, markedly broadening the physiological flow range [53, 54]. Accurate prediction of flow fields under varying inlet conditions is therefore essential for characterizing functional regulation and associated hemodynamic responses. However, such multi-condition variability substantially enlarges the underlying solution manifold, placing significant demands on PINN-based frameworks in terms of training efficiency, convergence robustness, and cross-condition generalization. Furthermore, in patients with severe stenosis, local Reynolds numbers may approach $10^3$, leading to increasingly complex flow structures, which further exacerbate the difficulty of accurately resolving coronary hemodynamics using PINNs. To date, unsupervised PINNs have shown limited success in patient-specific 3D coronary flow simulations with clinically acceptable efficiency, hindering clinical translation. Developing models that efficiently adapt to varying physiological conditions while preserving consistent flow physics remains challenging in patient-specific, angiography-derived three-dimensional coronary hemodynamic modeling.

Building on these considerations, we propose physics-informed hemodynamics modeling, an integrated deep learning framework for 3D hemodynamic analysis of stenotic coronary arteries from dual-view angiography, as shown in Figure 1d. The proposed framework enables efficient, minimally invasive coronary functional assessment, achieving strong agreement with CFD and clinical FFR across geometry, hemodynamics, and functional metrics, with an end-to-end runtime under 20 minutes.

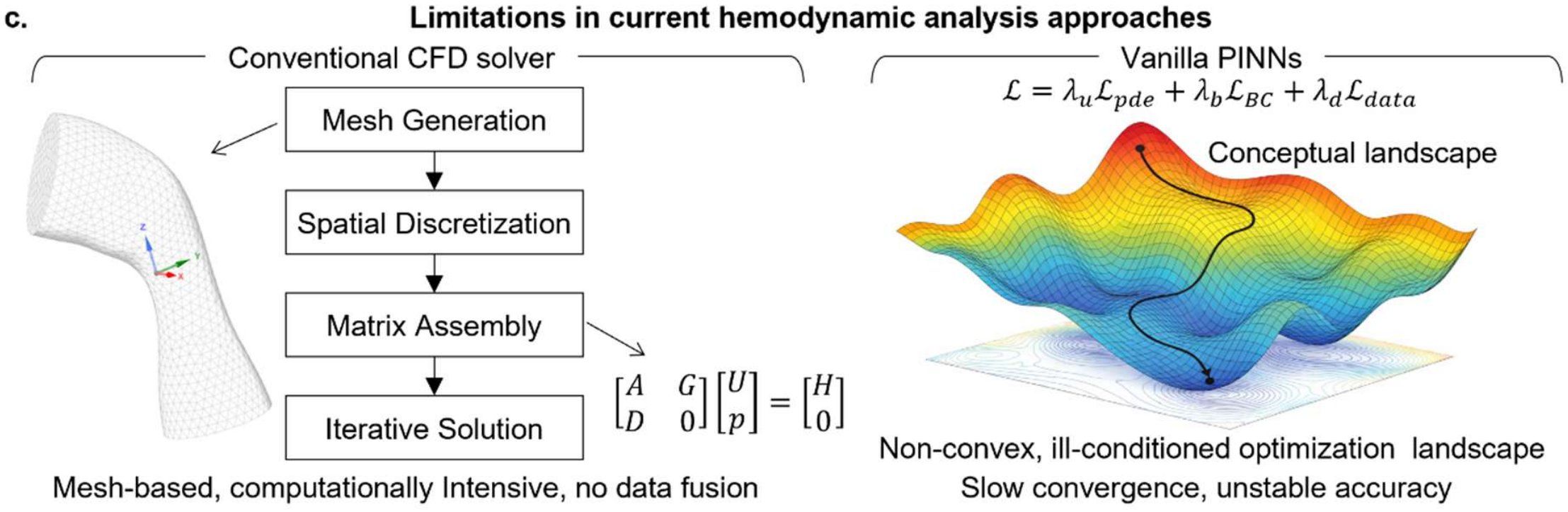

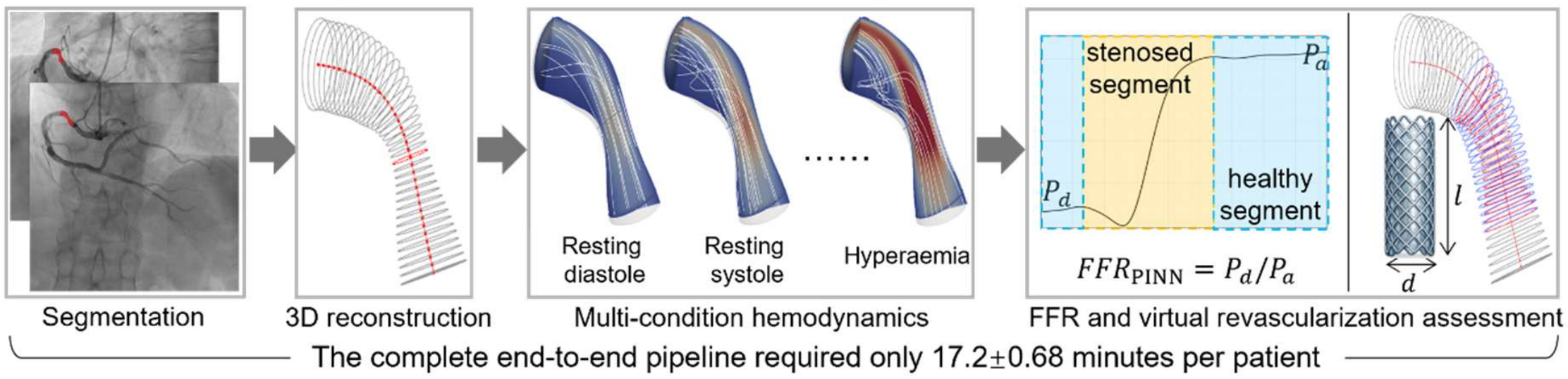

**Figure 1. Overview of current coronary stenosis assessment and the proposed physics-informed deep learning framework for hemodynamic analysis.** (a) Coronary stenosis caused by atherosclerotic plaque buildup, illustrated with invasive pressure-wire measurements used to compute fractional flow reserve (FFR) as the ratio of distal coronary pressure to aortic pressure. (b) Coronary angiography setup, showing C-arm imaging with common right coronary artery (RCA) view angles and the resulting two-dimensional projections. (c) Limitations in current hemodynamic analysis approaches. CFD-based coronary modeling requires volumetric meshing and fully prescribed boundary conditions, which are often unavailable clinically. Vanilla PINNs are mesh-free, but the coupled optimization of PDE and boundary residuals can be sensitive to loss balancing and initialization, particularly in complex vascular geometries. (d) End-to-end pipeline of the proposed framework. Manually segmented dual-view angiograms are used to reconstruct the patient-specific three-dimensional coronary centerline and radius profile, followed by multi-condition hemodynamic modeling, FFR estimation, and proof-of-concept virtual stent assessment. The complete workflow processes a patient case in approximately 17.2±0.68 minutes.

## 2. Overall Framework

The proposed framework consists of two components: (1) A 3D geometric reconstruction module based on dual-view 2D coronary angiography, which recovers patient-specific coronary centerlines and the corresponding along-vessel radius distribution (Figure 2a). (2) A hard-constrained blood flow solver based on an integrated JacobiNet–PINN architecture, which predicts velocity and pressure fields over the reconstructed geometry (Figure 2b).

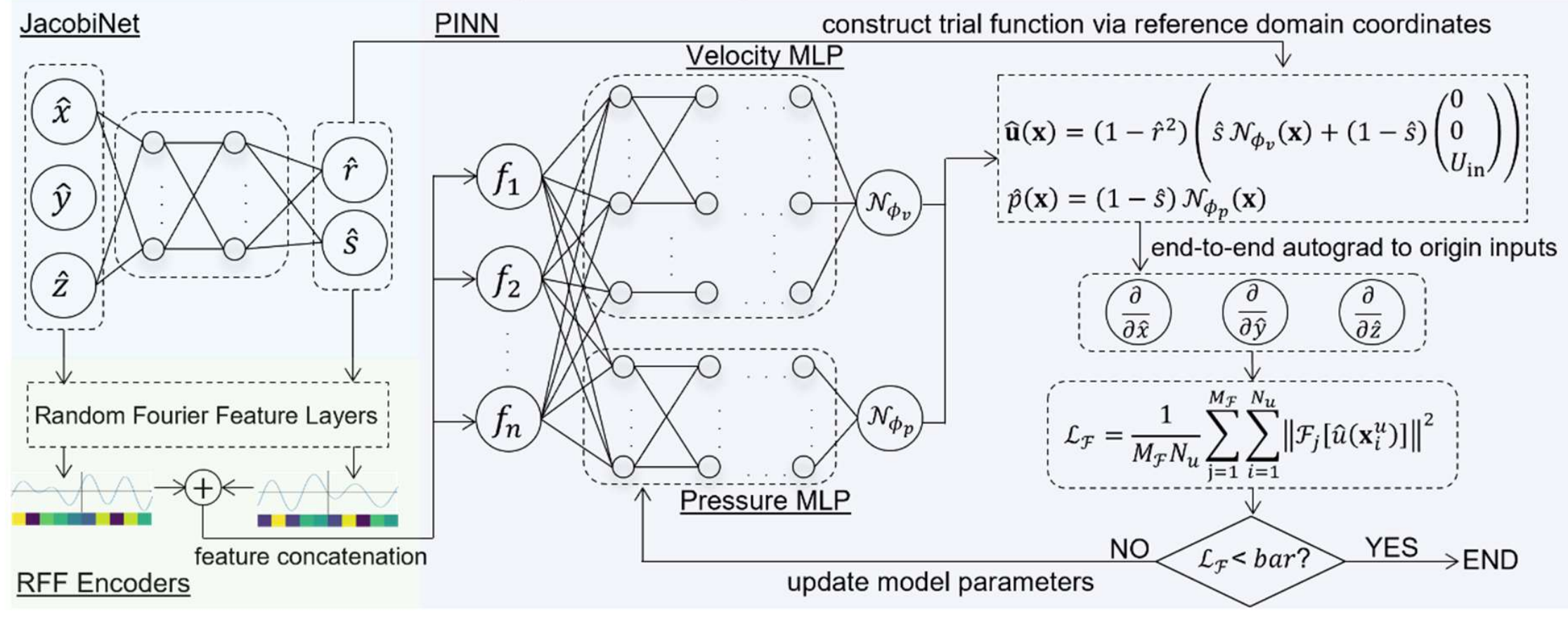

**Figure 2. Architecture of the proposed learning-based framework for coronary geometry reconstruction and hemodynamic modeling.** (a) Attention-enhanced convolutional neural network for 3D coronary geometry reconstruction. Two segmented angiographic views are represented as relative distance-transform maps and processed by a shared-weight ResNet-50 encoder. In the multiscale CBAM (MS-CBAM) pathway, CBAM recalibrates the deepest encoder features, while spatially max-pooled features from multiple encoder scales provide residual refinement of the radius prediction. The extracted features are fed into two multi-layer perceptrons (MLPs) to separately regress the 3D coronary centerline coordinates and the vessel radius, yielding a recovered centerline with associated radii and the corresponding sampled 3D point cloud representation. (b) JacobiNet–PINN framework for patient-specific 3D coronary flow modeling. A differentiable coordinate transformation network (JacobiNet) maps physical coordinates to reference-domain coordinates, which are embedded using random Fourier feature (RFF) encoders and concatenated with physical features. The physics-informed neural network (PINN) predicts velocity and pressure fields by constructing trial functions that satisfy boundary constraints, with end-to-end automatic differentiation used to enforce governing equations through physics-based loss terms. Model parameters are iteratively updated until convergence.

### 2.1 Attention-enhanced CNN for Coronary Geometry Reconstruction

For each stenotic segment, two coronary angiographic images from distinct projection angles with clear vessel opacification are selected and segmented. For each projection, the nominal object-plane pixel spacing was harmonized using the DICOM-recorded imager pixel spacing (IPS), source-to-object distance (SOD), and source-to-image distance (SID):

$$p_{\mathrm{obj}} = \mathrm{IPS}\frac{\mathrm{SOD}}{\mathrm{SID}}. \tag{1}$$

Each clinical projection was then resampled once to match the object-plane pixel spacing used for synthetic training:

$$k = \frac{p_{\mathrm{obj},c}}{p_{\mathrm{obj},t}} = \frac{\mathrm{IPS}_c(\mathrm{SOD}_c/\mathrm{SID}_c)}{\mathrm{IPS}_t(\mathrm{SOD}_t/\mathrm{SID}_t)}, \tag{2}$$

where the subscripts $c$ and $t$ denote the clinical and training configurations, respectively, with $\mathrm{IPS}_t = 0.258$ mm/pixel, $\mathrm{SID}_t = 1100$ mm, and $\mathrm{SOD}_t = 765$ mm.

A Euclidean distance transform was subsequently applied to enhance vascular structures and suppress background noise. The scale-harmonized images were converted to grayscale, centrally padded or cropped to exactly $128 \times 128$, and normalized to $[0,1]$ by division by 255. The two views are stacked along the view dimension to form the input tensor

$$\mathbf{X} = \mathrm{stack}\left(\left\{\mathbf{I}^{(v)}\right\}_{v=1}^{2}\right) \in \mathbb{R}^{2\times1\times H\times W}, \tag{3}$$

where $v \in \{1,2\}$ indexes the projection view, $\mathbf{I}^{(v)} \in \mathbb{R}^{1\times H\times W}$ denotes the corresponding single-channel projection image, and $H = W = 128$. The two views are processed by a shared-weight convolutional encoder based on ResNet50 [55] to extract vascular features.

To enhance sensitivity to elongated vessels and localized stenoses, we introduce a multiscale convolutional block attention module (MS-CBAM) into the radius-estimation pathway. Its deepest-feature recalibration is based on CBAM [56]. Given the deepest encoder output

$$\mathbf{F} \in \mathbb{R}^{C\times H'\times W'}, \tag{4}$$

where $C = 2048$ denotes the number of feature channels, and $H' = W' = 4$ represent the spatial height and width of the feature map, respectively. For each view, residual CBAM sequentially applies channel and spatial attention while preserving the original features through residual connections:

$$\mathbf{F}_c = \mathbf{F} + \mathbf{F} \odot M_c(\mathbf{F}), \tag{5}$$

$$\tilde{\mathbf{F}} = \mathbf{F}_c + \mathbf{F}_c \odot M_s(\mathbf{F}_c), \tag{6}$$

where $\odot$ denotes element-wise multiplication. The channel attention $\mathrm{M}_c \in \mathbb{R}^{C\times1\times1}$ is derived from global pooling, while the spatial attention $\mathrm{M}_s \in \mathbb{R}^{1\times H'\times W'}$ is computed from channel-aggregated features. The recalibrated deep representation $\tilde{\mathbf{F}}$ then serves as the semantic reference for multiscale aggregation. Features from the preceding encoder scales are spatially max-pooled to a common resolution and fused with $\tilde{\mathbf{F}}$, after which a residual mapping refines the predicted radius profile. This mechanism highlights geometry-relevant regions, including stenotic and highly curved segments, while suppressing responses from projection overlap and background interference.

For each view $v \in \{1,2\}$, the encoder backbone outputs a global feature vector $\mathbf{f}^{(v)} \in \mathbb{R}^{2048}$. The two feature vectors are concatenated to form a fused dual-view representation,

$$\mathbf{f}_{\text{fusion}} = [\mathbf{f}^{(1)}, \mathbf{f}^{(2)}] \in \mathbb{R}^{4096}, \tag{7}$$

which serves as the shared input to a multi-head regression architecture. Specifically, the centerline regression branch maps the fused features to a set of discrete 3D points $\widehat{\mathbf{X}} = \{\widehat{\mathbf{X}}_i\}_{i=1}^{N} \subset \mathbb{R}^3$ using a multilayer perceptron (MLP), where $N$ denotes the number of uniformly sampled centerline points along the vessel axis. In parallel, the radius regression branch takes the same fused features as input and predicts the corresponding vessel radii $\widehat{\mathbf{R}} = \{\widehat{R}_i\}_{i=1}^{N}$, which are defined pointwise along the reconstructed centerline.

To improve training stability and mitigate gradient interference between tasks, a stage-wise training strategy is employed. In the first stage, the backbone network and the centerline regression branch are optimized while the radius regression branch is frozen. In the second stage, the backbone network and the centerline branch are frozen, and only the radius regression branch is trained. The corresponding optimization objectives are defined using the mean squared error (MSE) losses:

$$\mathcal{L}_{\text{center}} = \frac{1}{N}\sum_{i=1}^{N} \| \widehat{\mathbf{X}}_i - \mathbf{X}_i^{\text{gt}} \|_2^2, \tag{8}$$

$$\mathcal{L}_{\text{radius}} = \frac{1}{N}\sum_{i=1}^{N} \left(\widehat{R}_i - R_i^{\text{gt}}\right)^2, \tag{9}$$

Here, $\widehat{\mathbf{X}}_i$ and $\mathbf{X}_i^{\text{gt}}$ denote the predicted and ground-truth centerline points, respectively, while $\widehat{R}_i$ and $R_i^{\text{gt}}$ represent the corresponding predicted and ground-truth vessel radii at the $i$-th centerline location.

The predicted discrete centerline points are ordered using a nearest-neighbor criterion, smoothed with one-dimensional Gaussian filtering, and densified by arc-length resampling to achieve approximately uniform spacing. Using the predicted radii, circular cross-sections are generated along the centerline and lofted to construct a continuous, smooth, and topologically consistent 3D vessel lumen surface. The reconstructed geometry is finally exported in point-cloud CSV and the standard STL (STereoLithography) surface format, enabling mesh-free PINN simulations and mesh-based CFD validation, respectively.

### 2.2 JacobiNet-PINN for 3D Coronary Flow Modeling

Based on the reconstructed geometry, a learning-based coordinate transformation, termed JacobiNet [57], maps physical Cartesian coordinates to normalized radial–axial parametric coordinates, embedding geometry-aware priors into the PINN framework for complex vascular domains. This formulation enables the solution of 3D steady-state blood flow on a regular reference domain implicitly defined by patient-specific geometry. Moreover, by employing trial functions constructed in the reference domain, boundary conditions are analytically embedded into the network architecture. This physics-guided design also allows efficient adaptation to varying inlet flow conditions through limited adjustment of internal degrees of freedom, enabling stable and efficient cross-condition transfer learning.

### 2.2.1 Geometry-based Cylindrical Parameterization

Given the predicted discrete centerline point set, the centerline and its corresponding vessel radius are represented as discrete parametric functions of the arc-length parameter, i.e.,

$$\hat{\mathbf{X}}_i = \hat{\mathbf{X}}(\hat{l}_i), \hat{R}_i = \hat{R}(\hat{l}_i), i = 1,2, \dots, N, \tag{10}$$

where $\hat{l}_i$ denotes the accumulated arc length along the centerline, starting from the inlet $\hat{l}_1 = 0$ and ending at the outlet $\hat{l}_N = L$. Here, $N$ represents the number of discrete centerline samples and $L$ is the total arc length of the centerline.

At each centerline location, a local orthonormal basis $\{\mathbf{t}(\hat{l}), \mathbf{n}(\hat{l}), \mathbf{b}(\hat{l})\}$ is constructed using a parallel transport frame, where $\mathbf{t}(\hat{l})$ is the unit tangent vector, and $\mathbf{n}(\hat{l})$ and $\mathbf{b}(\hat{l})$ denote the normal and binormal vectors, respectively.

Based on the predicted radius function $\hat{R}(\hat{l})$, circular cross-sectional point clouds are generated at each centerline location. Accordingly, any point $\hat{\mathbf{x}} \in \Omega$ within the vascular domain can be expressed in the local coordinate system as

$$\hat{\mathbf{x}} = \hat{\mathbf{X}}(\hat{l}) + \hat{\rho}(\cos\hat{\theta}\, \mathbf{n}(\hat{l}) + \sin\hat{\theta}\, \mathbf{b}(\hat{l})), \tag{11}$$

where $\hat{\rho} \in [0, \hat{R}(\hat{l})]$ denotes the radial distance from the centerline and $\hat{\theta} \in [0,2\pi)$ is the angular coordinate. Furthermore, normalized axial and radial coordinates are introduced as

$$\hat{s} = \frac{\hat{l}}{L} \in [0,1], \hat{r} = \frac{\hat{\rho}}{\hat{R}(\hat{l})} \in [0,1]. \tag{12}$$

Through this geometric parameterization, discrete sample points within the vascular domain are mapped from physical Cartesian coordinates $(\hat{x}, \hat{y}, \hat{z})$to the parametric space $(\hat{s}, \hat{\theta}, \hat{r})$.

### 2.2.2 JacobiNet-based Differentiable Coordinate Transformation

For partial differential equations (PDE) solved in transformed reference domains, access to the Jacobian matrix of the reference coordinates with respect to the physical coordinates is required to support chain-rule evaluation of differential operators such as gradients, divergence, and the Laplacian,

$$\mathrm{J}(\mathrm{x}) = \frac{\partial(\hat{s},\hat{\theta},\hat{r})}{\partial(\hat{x},\hat{y},\hat{z})}. \tag{13}$$

Since the Jacobian matrix of the geometric parameterization defined in Section 2.2.1 is not analytically available, we introduce JacobiNet to learn a continuous and differentiable coordinate transformation operator. Specifically, JacobiNet is formulated as a shallow fully connected neural network that approximates the mapping below,

$$\Phi_\vartheta: \hat{\mathbf{x}} = (\hat{x}, \hat{y}, \hat{z}) \in \mathbb{R}^3 \;\; \longrightarrow \;\; \hat{\boldsymbol{\xi}} = (\hat{s}, \hat{\theta}, \hat{r}) \in \mathbb{R}^3, \tag{14}$$

where $\vartheta$ denotes the network parameters.

The training of JacobiNet is supervised using samples generated from the geometric parameterization process, including both interior and boundary points of the vascular domain. Let the interior point set be denoted as $\{(\hat{\mathbf{x}}_i^{\mathrm{in}}, \hat{\boldsymbol{\xi}}_i^{\mathrm{in}})\}_{i=1}^{N_{\mathrm{in}}}$, and the boundary point set as $\{(\hat{\mathbf{x}}_i^{\mathrm{bd}}, \hat{\boldsymbol{\xi}}_i^{\mathrm{bd}})\}_{i=1}^{N_{\mathrm{bd}}}$, where $\hat{\boldsymbol{\xi}}_i = (\hat{s}_i, \hat{\theta}_i, \hat{r}_i)$ represents the reference-domain coordinates corresponding to the physical coordinates $\hat{\mathbf{x}} = (\hat{x}_i, \hat{y}_i, \hat{z}_i)$. The training objective of JacobiNet is defined by minimizing the mean squared error between the predicted parametric coordinates and those obtained from geometric construction,

$$\mathcal{L}_{\mathrm{Jacobi}} = \frac{1}{N_{\mathrm{in}}} \sum_{i=1}^{N_{\mathrm{in}}} \| \mathbf{\Phi}_{\boldsymbol{\vartheta}}(\hat{\mathbf{x}}_i^{\mathrm{in}}) - \hat{\boldsymbol{\xi}}_i^{\mathrm{in}} \|_2^2 + \lambda \frac{1}{N_{\mathrm{bd}}} \sum_{i=1}^{N_{\mathrm{bd}}} \| \mathbf{\Phi}_{\boldsymbol{\vartheta}}(\hat{\mathbf{x}}_i^{\mathrm{bd}}) - \hat{\boldsymbol{\xi}}_i^{\mathrm{bd}} \|_2^2, \tag{15}$$

where $\lambda$ is a weighting parameter that balances the contributions of interior and boundary supervision. In this study, the boundary-supervision weight is fixed at $\lambda = 10$, adopted from our previous JacobiNet study [57].

Through the learned mapping $\mathbf{\Phi}_{\boldsymbol{\vartheta}}$ provided by JacobiNet, the transformation between physical and geometry-parameterized coordinates is embedded as an end-to-end differentiable module within the PINN computational graph. Automatic differentiation enables efficient computation of first- and higher-order derivatives. As a result, coordinate transformations and their effects on physical quantities can be handled in a unified manner, without explicit Jacobian construction or manual chain-rule reformulation of the governing equations. The underlying concept and its advantages in numerical stability and computational efficiency have been systematically analyzed in our previous work [57]. Accordingly, JacobiNet adopts the same network architecture, loss formulation, and training strategy.

For the Navier–Stokes equations defined in coronary geometries, we observe that the angular coordinate $\hat{\theta}$ does not contribute to the construction of hard constraints in the PINN formulation (see Eqs. 27, 28). Consequently, JacobiNet only needs to learn a reduced geometry-aware coordinate mapping, $\mathbf{\Phi}_{\boldsymbol{\vartheta}}$: $(\hat{x}, \hat{y}, \hat{z}) \in \mathbb{R}^3 \longrightarrow (\hat{s}, \hat{r}) \in \mathbb{R}^2$.

### 2.2.3 PINN Formulation with Velocity–Pressure Decoupled Networks

A PINN typically employs a fully connected feedforward architecture $\mathcal{N}: \mathbb{R}^{D_i} \to \mathbb{R}^{D_o}$, where the computation at the $k$-th layer is defined as

$$\mathcal{N}^k(\mathrm{x}) = \Phi(\mathrm{W}_k \mathcal{N}^{k-1}(\mathrm{x}) + \mathrm{b}_k), 1 \le k \le L-1, \tag{16}$$

with activation function $\Phi(\cdot)$, trainable parameters $\{\mathbf{W}_k, \mathbf{b}_k\}$, and input layer $\mathcal{N}^0(\mathbf{x}) = \mathbf{x}$.

Training is performed by minimizing a composite loss function consisting of PDE residuals, boundary condition constraints, and data supervision,

$$\mathcal{L}_{\mathrm{total}} = \lambda_u \mathcal{L}_u + \lambda_b \mathcal{L}_b + \lambda_d \mathcal{L}_d, \tag{17}$$

where

$$\mathcal{L}_u = \frac{1}{M_u} \frac{1}{N_u} \sum_{j=1}^{M_u} \sum_{i=1}^{N_u} \left\| \mathcal{F}_j[\hat{\mathbf{u}}(\mathbf{x}_i^u)] \right\|^2, \tag{18}$$

$$\mathcal{L}_b = \frac{1}{M_b}\frac{1}{N_b}\sum_{j=1}^{M_b}\sum_{i=1}^{N_b}\left\|\mathcal{B}_j\left[\hat{\mathbf{u}}\left(\mathbf{x}_i^b\right)\right]\right\|^2, \tag{19}$$

$$\mathcal{L}_d = \frac{1}{N_d}\sum_{i=1}^{N_d}\left\|\hat{\mathbf{u}}\left(\mathbf{x}_i^d\right) - \tilde{\mathbf{u}}\left(\mathbf{x}_i^d\right)\right\|^2. \tag{20}$$

Here, $\{\mathbf{x}_i^u\}_{i=1}^{N_u}$, $\{\mathbf{x}_i^b\}_{i=1}^{N_b}$, and $\{(\mathbf{x}_i^d, \tilde{\mathbf{u}}(\mathbf{x}_i^d))\}_{i=1}^{N_d}$ denote collocation points for enforcing PDE residuals, boundary conditions, and data supervision, respectively. The quantities $M_u$ and $M_b$ represent the numbers of governing PDE and boundary operators, respectively. Specifically, $\mathcal{F}_j$ denotes the $j$th interior governing-equation residual operator, comprising the 3D steady Navier–Stokes momentum residuals and the incompressibility residual. $\mathcal{B}_j$ denotes the $j$th boundary-condition residual operator associated with the inlet, outlet, or vessel wall. $\hat{\mathbf{u}}$ denotes the network prediction, while $\tilde{\mathbf{u}}$ is the labeled ground-truth value. $\lambda_u$, $\lambda_b$, and $\lambda_d$ control the contributions of each loss component during training. In the label-free forward calculations of our study, we set $\lambda_u = 1$, $\lambda_d = 0$, and $\lambda_b = 0$ (boundary conditions were imposed analytically). In the sparse-data assimilation experiments (Section 5), we used $\lambda_d = 0.001$ for the simulated Doppler observations, and $\lambda_d = 0.01$ for the simulated MRI observations.

At the coronary scale, blood can be reasonably modelled as an incompressible Newtonian fluid [58, 59]. Its steady 3D flow is governed by the dimensionless Navier–Stokes equations,

$$\mathbf{u}\cdot\nabla\mathbf{u} + \nabla p - \frac{1}{\mathrm{Re}}\nabla^2\mathbf{u} = \mathbf{0}, \tag{21}$$

$$\nabla\cdot\mathbf{u} = \mathbf{0}, \tag{22}$$

where $\mathbf{u} = (u, v, w)$ denotes the velocity vector, $p$ is the pressure, and Re is the Reynolds number. The blood density and dynamic viscosity are taken as $\rho = 1060\ \mathrm{kg/m^3}$ and $\mu = 0.004\ \mathrm{Pa\cdot s}$, respectively [60, 61]. The vascular domain is subject to a parabolic inlet velocity profile, a zero-pressure outlet condition, and no-slip wall boundaries, defined as

$$\mathbf{u} = U_{\max}\left(1 - \frac{r^2}{\hat{R}_0{}^2}\right), \text{ on } \Gamma_{\text{inlet}}, \tag{23}$$

$$p = p_{\text{outlet}}, \text{ on } \Gamma_{\text{outlet}}, \tag{24}$$

$$\mathbf{u} = \mathbf{0}, \text{ on } \Gamma_{\text{wall}}, \tag{25}$$

where $r$ is the radial distance from the centerline and $\hat{R}_0$ is the inlet radius. For each case, the geometry-adaptive reference scales were defined as $L_0 = R_{\text{inlet}}/2$, $U_0 = U_{\text{mean}}(R_{\text{inlet}}/R_{\min})^2$, and $P_0 = \rho U_0^2$, where $U_{\text{mean}} = U_{\text{peak}}/2$. Accordingly, $\mathbf{x}^* = \mathbf{x}/L_0$, $\mathbf{u}^* = \mathbf{u}/U_0$, and $p^* = p/P_0$, yielding $Re = U_0 L_0/\nu$ in Eq. (21). Boundary data and predicted fields were transformed consistently using these scales.

To enhance the efficiency of physical constraint propagation in complex domains, geometry-parameterized coordinates $(\hat{s}, \hat{r}) \in \mathbb{R}^2$ and physical Cartesian coordinates $(\hat{x}, \hat{y}, \hat{z}) \in \mathbb{R}^3$ are fused through Random Fourier Feature (RFF) encoder layers.

Specifically, the input features are first mapped into a fixed random Fourier basis space,

$$\boldsymbol{\phi}(\mathbf{x}) = [\cos(2\pi\mathbf{B}\mathbf{x}), \sin(2\pi\mathbf{B}\mathbf{x})], \tag{26}$$

where $\mathbf{B} \in \mathbb{R}^{D/2\times d}$ is a random projection matrix sampled from a zero-mean Gaussian distribution, with its variance controlled by the hyperparameter $\sigma$. Prior studies have shown that Random Fourier Features enhance the representation of high-frequency components, thereby mitigating the spectral bias commonly observed in conventional MLPs when modeling high-frequency physical fields [62-64].

Unlike unified spectral encoding of all inputs, the proposed spectral decoupling strategy assigns distinct roles to geometry-parameterized and physical coordinates in the spectral domain. The geometry coordinates $(\hat{r}, \hat{s})$ capture global axial flow evolution and radial structural characteristics, providing stable geometric constraints, while the physical coordinates $(\hat{x}, \hat{y}, \hat{z})$ model localized spatial variations of velocity and pressure. This complementary spectral representation enables the network to simultaneously capture global geometry and local flow physics within a unified framework.

Subsequently, the network adopts a velocity–pressure decoupled subnetwork architecture, employing two separate fully connected feedforward networks, $\mathcal{N}_{\phi_v}$ and $\mathcal{N}_{\phi_p}$, to predict the velocity and pressure fields, respectively. This design is motivated by the pronounced asymmetry between velocity and pressure variables in incompressible Navier–Stokes equations in terms of learning difficulty and physical roles. Compared with architectures that share a common backbone, the decoupled structure effectively alleviates gradient competition, leading to more stable training and more physically consistent solutions.

#### 2.2.4 Hard-Constrained Trial Functions and Efficient Parameter Transfer

The computational domain is first rigidly transformed via rotation and translation such that the inlet plane is perpendicular to the $z$-axis and the inlet velocity is aligned with the positive $z$-direction.

Following established trial-function approaches for the exact imposition of boundary conditions in neural PDE solvers [65, 66], we analytically embed the prescribed inlet-velocity, no-slip wall, and outlet-pressure conditions into the network outputs. In the normalized radial–axial coordinate system $(\hat{r}, \hat{s}) \in [0,1]^2$, the corresponding hard-constrained trial functions are defined as

$$\hat{\mathbf{u}}(\mathbf{x}) = (1-\hat{r}^2)\left(\hat{s}\,\mathcal{N}_{\phi_v}(\mathbf{z}) + (1-\hat{s})\begin{pmatrix}0\\0\\U_{\text{in}}\end{pmatrix}\right), \tag{27}$$

$$\hat{p}(\mathbf{x}) = (1-\hat{s})\,\mathcal{N}_{\phi_p}(\mathbf{z}), \tag{28}$$

where $\mathcal{N}_{\boldsymbol{\phi}_v}: \mathbb{R}^D \to \mathbb{R}^3$ and $\mathcal{N}_{\boldsymbol{\phi}_p}: \mathbb{R}^D \to \mathbb{R}$ are the velocity and pressure subnetworks operating on the RFF-encoded feature space, respectively. The augmented input vector $\mathbf{z} = [\,\boldsymbol{\phi}(\hat{x}, \hat{y}, \hat{z}),\ \boldsymbol{\phi}(\hat{r}, \hat{s})\,]$ is formed by concatenating the Random Fourier Feature encodings of the physical coordinates and the normalized geometric parameters, and $U_{\text{in}}$ denotes the normalized inlet velocity magnitude.

The proposed trial function formulation analytically enforces boundary constraints at the vessel wall, inlet, and outlet by construction. At the vessel wall ($\hat{r} = 1$), the velocity prediction automatically satisfies $\hat{\mathbf{u}}(\mathbf{x}) = \mathbf{0}$, strictly enforcing the no-slip condition. Within the lumen ($\hat{r} < 1$), the formulation naturally induces a radial velocity prior consistent with laminar Poiseuille flow at low Reynolds numbers [58, 67, 68]. At the inlet ($\hat{s} = 0$), the axial velocity component exactly satisfies $\hat{w} = (1 - \hat{r}^2)U_{\text{in}}$, analytically imposing a parabolic inlet profile, while at the outlet ($\hat{s} = 1$) the pressure prediction reduces to $\hat{p}(\mathbf{x}) = 0$. By embedding boundary conditions directly into the network output, this approach eliminates the need for boundary penalty terms and avoids gradient instability arising from competing loss components, which has been shown to be exacerbated by complex boundary conditions in PINN optimization, as discussed in previous work [57].

Moreover, under the proposed hard-constrained solution template, different inlet flow conditions are distinguished solely by the parameter $U_{\text{in}}$. When transferring to new flow-rate conditions, the network architecture and geometric parameterization remain unchanged, requiring only a small number of optimization iterations to recover consistent flow patterns. By applying Random Fourier Features to both the reference-domain coordinates $(\hat{r}, \hat{s})$ and the physical coordinates $(\hat{x}, \hat{y}, \hat{z})$, the model preserves dominant flow structures and shear-layer morphology across varying Reynolds numbers. As a result, changes in inlet velocity primarily modulate flow magnitude and local shear scales rather than inducing global structural reorganization. As demonstrated in the subsequent experiments, this property substantially simplifies cross-condition transfer learning, enabling rapid convergence and significantly reducing redundant training costs for multi-velocity coronary functional assessment.

While our previous work [57] introduced a neural coordinate transformation strategy to enhance the stability of physics-informed neural networks in irregular domains, this study advances beyond geometric normalization by reframing patient-specific coronary flow modeling as a physics-informed, condition-transferable inference problem defined on a low-dimensional solution manifold. Specifically, the present study makes four specific advances: First, we developed an attention-enhanced CNN that reconstructs patient-specific three-dimensional right coronary artery geometries from two clinically acquired angiographic projections, thereby providing the anatomical basis for subsequent hemodynamic modeling. Second, we extended JacobiNet from parametrically generated vessel benchmarks [57] to angiography-derived, patient-specific three-dimensional coronary anatomies. Third, building on established RFF encoding methods, we separately mapped the natural and transformed coordinates into complementary Fourier feature spaces and subsequently fused their representations. This integration enables the network to capture both global vascular structure and localized flow dynamics. Finally, we integrated analytically enforced boundary constraints with embedded physical priors into the network architecture and coupled them with a transfer learning strategy, enabling rapid transfer across physiological flow conditions. Together, these advances mark a transition from PINN stabilization in idealized geometries (JacobiNet) toward clinically consistent hemodynamics modeling in realistic coronary anatomies.

# 3. Results on Synthetic Dataset

We first evaluated the proposed framework on a large-scale synthetic right coronary artery (RCA) dataset comprising 50,000 cases. Anatomically realistic 3D coronary geometries with statistically representative stenosis distributions were synthesized using an RCA-informed data generation pipeline [33]. Coronary CTA data from 10 patients were retrospectively collected and de-identified at the University of Michigan Hospital, with approval from the University of Michigan IRBMED (HUM00155491). CRIMSON was used to define patient-informed geometric ranges, from which synthetic centerlines and radius profiles were generated by random sampling, spline interpolation, and geometric augmentation. Stenoses were introduced by Gaussian-shaped modulation of the vessel radius along an equal–arc-length centerline, with location, length, and severity sampled from clinically reported distributions [69]. The resulting geometries were projected into dual-view coronary angiographic images using a calibrated cone-beam imaging model. The synthetic dataset was randomly split into training, validation, and test sets with a ratio of 80% / 10% / 10%, respectively. Geometric reconstruction was evaluated on the complete held-out test set of 5,000 cases, while hemodynamic evaluation was performed on a fixed 100-case cohort, generated from the 5,000-case test set through stratified random sampling without replacement across four prespecified reference stenosis bands. A detailed description of the synthetic data generation procedure and the corresponding parameter distributions is given in **Appendix-A**.

For reference-solution generation, reference CFD used ANSYS Fluent 2024 R1 (v24.1), three-dimensional double precision, and steady incompressible laminar Navier–Stokes equations. Blood density and viscosity were 1,060 kg/m³ and 0.004 Pa·s. Boundaries were a parabolic inlet, zero-gauge-pressure outlet, and rigid no-slip wall. Watertight Geometry meshes used curvature-based triangular surfaces (wall size 0.075 mm, growth 1.10, normal angle 20°), an unstructured tetrahedral core (maximum 0.600 mm, growth 1.20), and ten wedge-prism layers (first layer 0.010 mm, growth 1.10, total 0.159 mm). SIMPLEC was combined with least-squares cell-based gradients, second-order pressure, and second-order-upwind momentum. Simulations stopped at continuity $10^{-5}$ and momentum $10^{-7}$ or 2,000 iterations; acceptance required continuity $10^{-3}$, momentum $10^{-5}$. Five-level mesh-independence and case-wise GCI results are in **Appendix-B**.

All experiments were implemented using the PyTorch framework and conducted on an NVIDIA GeForce RTX 4090D GPU. In the 3D reconstruction module, the ResNet-50 backbone was initialized with ImageNet-pretrained weights [70, 71]. The proposed network contains approximately $2.90 \times 10^7$ trainable parameters. A two-stage training scheme was adopted, with the backbone and centerline branch trained for 100 epochs, followed by radius branch training for 50 epochs with the former frozen. For the hemodynamic modeling module, the proposed JacobiNet–PINN framework contains $1.33 \times 10^5$ trainable parameters. PINN training used fixed severity-dependent horizons of 10,000, 20,000, 30,000, and 40,000 optimizer steps for reconstructed-geometry stenosis bands of 0–24%, 25–49%, 50–69%, and 70–99%, respectively. Detailed descriptions of the network architectures, loss formulations, and training hyperparameters are provided in **Appendix-B**.

### 3.1 Evaluation Metrics

The geometric reconstruction accuracy was quantified using the root mean squared error (RMSE) for centerline coordinates and the mean absolute error (MAE) for vessel radius prediction:

$$\mathrm{RMSE}_{xyz} = \sqrt{\frac{1}{N}\sum_{i=1}^{N} \| \mathbf{X}_i^{\mathrm{pred}} - \mathbf{X}_i^{\mathrm{ref}} \|_2^2}, \tag{29}$$

$$\mathrm{MAE}_r = \frac{1}{N}\sum_{i=1}^{N} | r_i^{\mathrm{pred}} - r_i^{\mathrm{ref}} |, \tag{30}$$

where $\mathbf{X}_i^{\mathrm{pred}}$ and $\mathbf{X}_i^{\mathrm{ref}}$ denote the predicted and reference 3D centerline coordinates, and $r_i^{\mathrm{pred}}$ and $r_i^{\mathrm{ref}}$ denote the corresponding vessel radii at the $i$-th centerline location.

Because the clinical FFR estimate depends directly on the pressure loss across the stenotic segment, pressure-drop error was defined as the primary hemodynamic endpoint. For both PINN and CFD, the pressure drop was calculated using the area-weighted mean static pressures on the same inlet and outlet surfaces:

$$\Delta P = \bar{p}_{\mathrm{in}} - \bar{p}_{\mathrm{out}}. \tag{31}$$

The corresponding absolute and percentage errors were defined as:

$$E_{\Delta P} = |\Delta P_{\mathrm{PINN}} - \Delta P_{\mathrm{CFD}}|, \tag{32}$$

$$APE_{\Delta P} = \frac{|\Delta P_{\mathrm{PINN}} - \Delta P_{\mathrm{CFD}}|}{|\Delta P_{\mathrm{CFD}}|} \times 100\%. \tag{33}$$

Flow-field prediction accuracy was evaluated using the relative $L_2$ error, computed separately within the whole field, stenotic throat, and post-stenotic recirculation region,

$$L_2 = \frac{\|\mathbf{q}^{\mathrm{pred}} - \mathbf{q}^{\mathrm{ref}}\|_2}{\|\mathbf{q}^{\mathrm{ref}}\|_2}, \tag{34}$$

where $\mathbf{q}^{\mathrm{pred}}$ and $\mathbf{q}^{\mathrm{ref}}$ denote the predicted and reference flow variables (velocity or pressure), respectively. The reference solutions were obtained from high-fidelity CFD simulations.

### 3.2 Accuracy of Geometry Reconstruction

The centerline and radius branches converged at epochs 78 and 47, respectively, with validation losses decreasing rapidly and stabilizing thereafter. Quantitative evaluation of 3D geometric reconstruction was performed on the test set. Across the 5,000 test cases, the reconstruction error of the centerline coordinates was $\mathrm{RMSE}_{xyz} = 0.288 \pm 0.158$, while the vessel radius prediction error was $\mathrm{MAE}_r = 0.029 \pm 0.018$, with an overall four-grade error rate of 7.84%.

Table 1 further presents case-wise reconstruction error versus diameter stenosis. Across different stenosis severities, both centerline and radius reconstruction errors remain relatively stable and do not exhibit systematic divergence with increasing stenosis severity.

**Table 1. Severity-stratified 3D reconstruction errors for the 5,000-case synthetic test set.**

| Metric (mean ± SD) | All | 0–24% | 25–49% | 50–69% | 70–99% |
|---|---|---|---|---|---|
| Centerline RMSE, mm | 0.288 ± 0.158 | 0.295 ± 0.165 | 0.282 ± 0.153 | 0.290 ± 0.160 | 0.288 ± 0.157 |
| Radius MAE, mm | 0.029 ± 0.018 | 0.028 ± 0.018 | 0.028 ± 0.018 | 0.030 ± 0.017 | 0.032 ± 0.018 |

Linear regression analysis shows only weak correlations between reconstruction error and diameter stenosis. Pearson correlation analysis showed negligible associations between reconstruction error and reference diameter stenosis in the 5,000-case test set. The correlation was $r = -0.005$ for centerline RMSE and $r = 0.06$ for vessel-radius MAE. The small positive tendency in radius error may reflect the steeper axial radius gradients and reduced visible lumen area in highly stenotic regions, which provide sparser local cues in the 2D projections. Nevertheless, the effect size was minimal, and the stratified error distributions remained stable, supporting robust reconstruction performance across stenosis severities.

As shown in Figure 3a, to qualitatively assess model robustness, four representative synthetic cases spanning the full clinical CAD-RADS spectrum from normal vessels to severe disease were selected for visualization, deliberately incorporating challenging geometric and imaging conditions.

(i) Healthy or mild stenosis (0–24%) under normal imaging conditions.

(ii) Mild-to-moderate stenosis (25–49%) with the lesion located in a highly curved segment.

(iii) Moderate stenosis (50–69%), where the two angiographic views are acquired at closely aligned projection angles.

(iv) Severe stenosis (70–99%), under normal imaging conditions.

The stenosis severity is quantified by the diameter stenosis ratio, defined as

$$\text{Stenosis} = \left(1-\frac{D_{\min}}{D_{\text{in}}}\right) \times 100\%, \tag{35}$$

where $D_{\min}$ denotes the minimum lumen diameter at the lesion and $D_{\text{in}}$ represents the reference vessel diameter in the inlet of the stenosed segment.

Current assessment based solely on 2D angiography may substantially under- or overestimate stenosis severity. For example, in case (iii), angiographic evaluation from both views suggested <40% stenosis, whereas the true stenosis reached 52.3%. In case (iv), the stenosis severity was still underestimated from the 2D angiography, which may lead to misinterpretation of lesion severity in clinical decision-making. However, the proposed method consistently preserves the overall vascular morphology, stenosis location, and stenosis severity across all cases. The table in Figure 3b shows that the estimated stenosis rates differ from the ground truth by less than 2%. Furthermore, the reconstructed centerline and radius $[\hat{X}, \hat{Y}, \hat{Z}, \hat{R}]$, together with the low error distributions along the centerline points, demonstrate strong robustness and stability across different clinical grades and challenging imaging conditions.

On the 5,000-case benchmark, comparing with Iyer et al. [33] baseline, our model reduced centerline RMSE, radius MAE, and diameter-stenosis MAE by 19.9%, 62.1%, and 77.0%, respectively. Complete results and ablation studies are provided in **Appendix-C**.

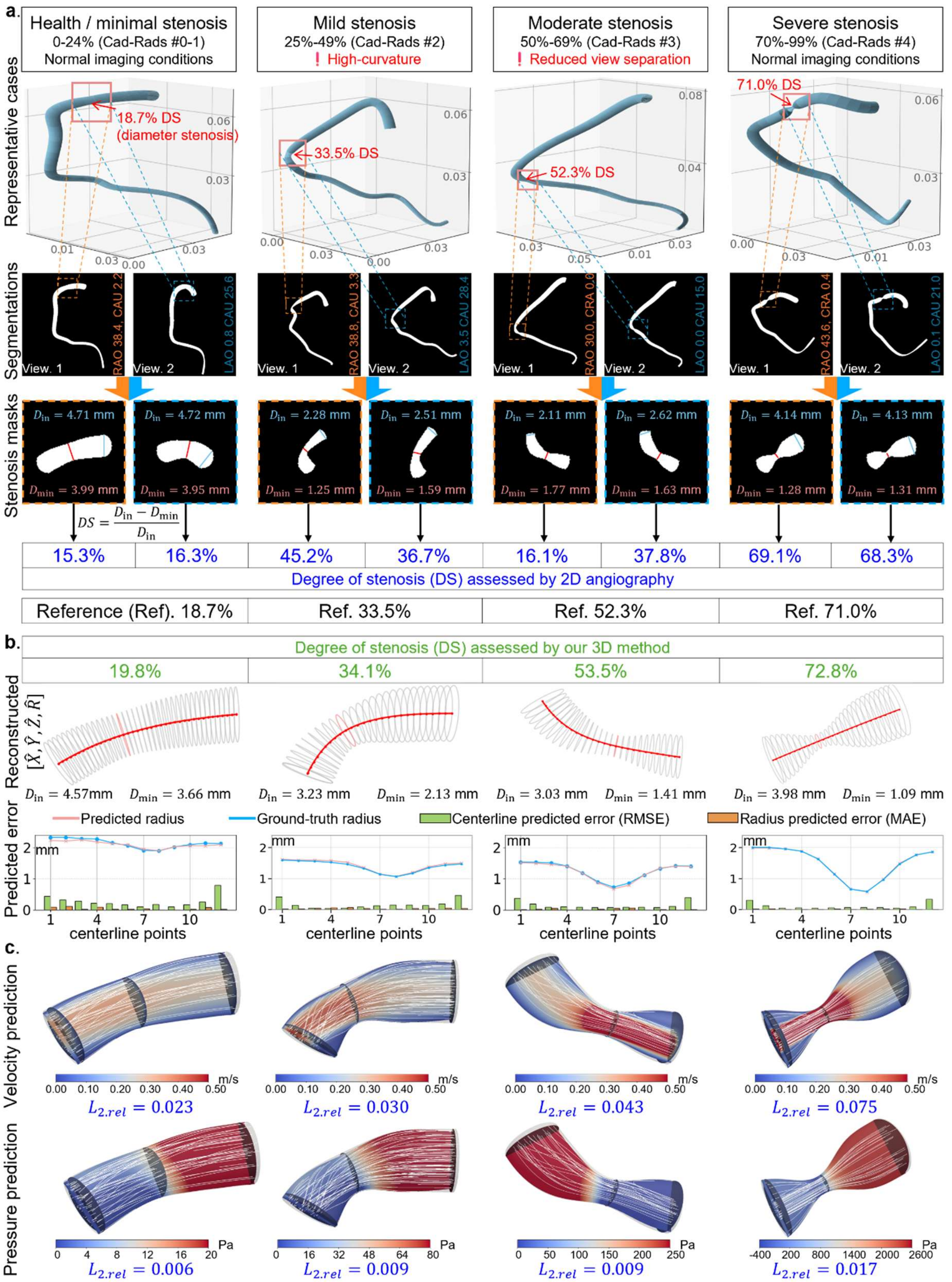

**Figure 3. Representative results of synthetic coronary geometry reconstruction and hemodynamic modeling under varying stenosis severity and imaging conditions.** (a) Example right coronary artery (RCA) cases spanning healthy/minimal, mild, moderate, and severe stenosis (CAD-RADS 0–4). Projections exhibit

challenging imaging conditions, including high-curvature segments, and limited view separation. Corresponding dual-view projections and acquisition angles are shown for each case. The bottom panel shows stenosis segmentation and diameter reduction assessment. As a projection-based approach, angiography may misestimate lesion severity relative to the reference vessel. (b) The proposed three-dimensional method enables more accurate and geometrically consistent quantification of lesion morphology. Geometry reconstruction results of the attention-enhanced CNN are presented, where the recovered 3D centerlines and vessel radii are visualized for representative cases. Quantitative comparisons between predicted and ground-truth centerline coordinates, as well as radius profiles along the vessel centerline, further demonstrate the reconstruction accuracy. (c) Flow modeling with PINN-JacobiNet. Relative $L2$ errors with respect to CFD ground truth demonstrate the accuracy of the JacobiNet–PINN framework, alongside predicted velocity streamlines and pressure fields for representative cases.

### 3.3 Performance of Hemodynamic Modeling

Hemodynamic performance was evaluated on a fixed subset of 100 cases drawn from the held-out test set. The results are summarized in Table 2.

**Table 2. Severity-stratified errors in pressure drop, regional flow fields, and WSS for the 100-casesynthetic dataset.**

| Metric (mean ± SD) | All | 0–24% | 25–49% | 50–69% | 70–99% |
|---|---|---|---|---|---|
| *Trans-stenotic pressure-drop error* | | | | | |
| Absolute error, Pa | 64.16 ± 168.24 | 0.26 ± 0.27 | 0.43 ± 0.37 | 11.68 ± 20.56 | 244.27 ± 266.83 |
| Absolute percentage error, % | 2.100 ± 4.190 | 0.460 ± 0.320 | 0.310 ± 0.200 | 1.010 ± 1.400 | 6.640 ± 6.450 |
| *Regional relative-$L_2$ error* | | | | | |
| Global velocity | 0.095 ± 0.105 | 0.027 ± 0.002 | 0.031 ± 0.004 | 0.084 ± 0.053 | 0.237 ± 0.111 |
| Global pressure | 0.024 ± 0.041 | 0.008 ± 0.002 | 0.007 ± 0.002 | 0.013 ± 0.013 | 0.067 ± 0.064 |
| Throat velocity | 0.046 ± 0.028 | 0.025 ± 0.002 | 0.027 ± 0.003 | 0.046 ± 0.012 | 0.087 ± 0.020 |
| Throat pressure | 0.028 ± 0.043 | 0.011 ± 0.003 | 0.013 ± 0.006 | 0.018 ± 0.012 | 0.071 ± 0.068 |
| Post-stenotic recirculation velocity [a] | 0.173 ± 0.153 | — | 0.040 ± 0.006 | 0.108 ± 0.073 | 0.306 ± 0.151 |
| Post-stenotic recirculation pressure [a] | 0.171 ± 0.137 | — | 0.062 ± 0.016 | 0.157 ± 0.095 | 0.242 ± 0.163 |
| *WSS error* | | | | | |
| Relative-$L_2$ | 0.034 ± 0.039 | 0.014 ± 0.005 | 0.014 ± 0.005 | 0.033 ± 0.032 | 0.077 ± 0.049 |
| Low-WSS area error, pp [b] | 0.425 ± 0.806 | 0.032 ± 0.069 | 0.296 ± 0.540 | 0.871 ± 1.229 | 0.499 ± 0.647 |
| High-WSS area error, pp [b] | 1.274 ± 2.106 | 0.180 ± 0.260 | 0.300 ± 0.682 | 1.360 ± 1.757 | 3.257 ± 2.834 |

[a] Post-stenotic results include only CFD detected recirculation cases (n = 63 overall; n = 0, 13, 25, and 25 across the four bands);

[b] Low/high WSS thresholds are <4 and >40 dyne/cm².

For pressure-drop errors relative to CFD, the overall mean absolute percentage errors were 2.10 ± 4.19% in the 100-case synthetic cohort, increasing to 6.64 ± 6.45% for severe stenoses (70–99%). Nevertheless, among the 5 cases with pressure-drop percentage errors exceeding 10%, the CFD- and PINN-derived FFR classifications at the clinical threshold of 0.80 agreed in all cases. Thus, this result indicates that although severe stenosis posed greater challenges for flow prediction, the resulting errors did not alter the threshold-based FFR classification in these cases.

The results also show that regional errors remained low at the stenotic throat (velocity relative-$L_2$: 0.046 ± 0.028; pressure relative-$L_2$: 0.028 ± 0.043), where the mean CFD pressure loss was 1207.741 ± 1675.525 Pa, corresponding to 107.625% of the net inlet-to-outlet pressure drop; the value above 100% reflects subsequent pressure recovery. Errors were higher in the post-stenotic region of recirculation-positive cases (0.173 ± 0.153 and 0.171 ± 0.137, respectively); however, the mean absolute pressure change in this region was only 176.635 ± 188.056 Pa, equivalent to 14.63% of the mean throat pressure loss, and therefore had limited influence on FFR estimation.

Further, WSS predictions agreed closely with CFD, with an overall relative-$L_2$ error of 0.034 ± 0.039 and low- and high-WSS area errors of 0.425 ± 0.806 and 1.274 ± 2.106 percentage points, respectively. Even in the 70–99% stenosis group, the mean relative-$L_2$ error remained below 0.1, supporting the robustness of the area-based WSS analysis. To evaluate the contribution of individual components, ablation studies were conducted in **Appendix-C**.

# 4. Validations on Clinical Records

To further evaluate the effectiveness of the proposed method in real clinical settings, a retrospective analysis was conducted on coronary angiography data from 177 patients collected at West China Hospital of Sichuan University (Chengdu, China). The dataset comprised patients who underwent FFR intervention in the cardiac catheterization laboratory between May–October 2024. This study was conducted in accordance with the principles of the Declaration of Helsinki and was approved by the Ethics Committee of West China Hospital, Sichuan University. All data were anonymized prior to analysis, and the requirement for informed consent was waived due to the retrospective nature of the study.

Inclusion criteria were as follows: (i) lesions located in the right coronary artery (RCA); (ii) single-vessel coronary artery disease; (iii) availability of high-quality coronary angiography images with clearly visible target lesions; (iv) angiographically confirmed coronary artery stenosis; and (v) availability of corresponding invasive FFR measurements for reference. Patients not meeting these criteria were excluded from the analysis.

Among the 177 potentially eligible patients, 118 patients were excluded because no RCA stenosis was identified, and 15 because invasive FFR was measured in a coronary vessel other than the RCA. Of the remaining 44 patients, 5 were excluded because of poor angiographic image quality, 4 because only a single-view projection was available, and 3 because a branch was present at the stenotic segment. Consequently, 32 patients were included in the final clinical validation cohort (Figure 5c). Detailed patient-level information for the clinical cohort is provided in **Appendix-D**.

Lesion-boundary selection, reconstruction-range determination, manual segmentation, three-dimensional reconstruction, and flow modeling were performed blinded to invasive FFR. The prespecified sequence was de-identified angiography → lesion and range definition → segmentation → reconstruction → JacobiNet–PINN calculation → prediction lock → unblinding and comparison with invasive FFR. No patient was excluded by invasive FFR, prediction error, or classification outcome. The second observer independently re-annotated all 32 cases without access to the first observer's contours, invasive FFR, or downstream results. Because invasive FFR was acquired before the retrospective index analysis, the index-test results were unavailable to the reference-standard assessors. The results of the interobserver agreement analysis are provided in **Appendix-F**.

Based on the reconstructed geometries $\left[\hat{X}, \hat{Y}, \hat{Z}, \hat{R}\right]$, all cases were sampled at uniform spatial intervals. The sampling spacing was set to $1 \times 10^{-4}$m for interior points and $5 \times 10^{-5}$m for boundary points. Across 32 geometries, they contained 20,849–161,710 points (mean, 57,118) and 10,646–41,483 points (mean, 19,694), respectively.

This study was reported in accordance with the Standards for Reporting Diagnostic Accuracy (STARD) 2015 guidelines, with the completed checklist provided in the Supplementary Materials. No formal a priori sample-size calculation was performed because this was a retrospective proof-of-concept study. The sample size was determined by the availability of patients who underwent invasive FFR measurement and had complete angiographic data during the predefined study period.

### 4.1 Efficient Parameter Transfer Learning across Multiple Velocity Conditions

Coronary blood flow exhibits substantial velocity variations driven by the cardiac cycle and physiological regulation. As shown in Figure 4b-1, the mean velocity in the right coronary artery under resting conditions is approximately 16.9 cm/s, with systolic and diastolic averages of 14.4 cm/s and 20.9 cm/s, respectively [52, 53]. During FFR assessment, adenosine-induced maximal hyperemia markedly reduces microvascular resistance (Figure 4b-2), resulting in a substantial increase in coronary flow velocity to an average of 38.3 cm/s [53, 54].

Leveraging hard-constrained trial functions with embedded Poiseuille-flow priors, the proposed framework enables efficient parameter transfer across varying physiological flow conditions. The model is first trained under the lowest inlet velocity (resting systolic flow) and subsequently transferred to other inlet conditions. At each target condition, the transferred model is physics-only fine-tuned for only 1,000 steps using the governing equations. Across the 32 patients (96 transfers in total), the mean end-to-end wall time was $13.79 \pm 2.07$ s per target-condition transfer, substantially less than training a new PINN or rerunning CFD from scratch.

After averaging the four condition-specific errors within each patient, the mean ± sample SD across the 32 patients was 0.054 ± 0.023 for velocity and 0.023 ± 0.016 for pressure. Further discussion of computational efficiency and comparisons with simple linear and quadratic scaling controls are provided in **Appendix-E**.

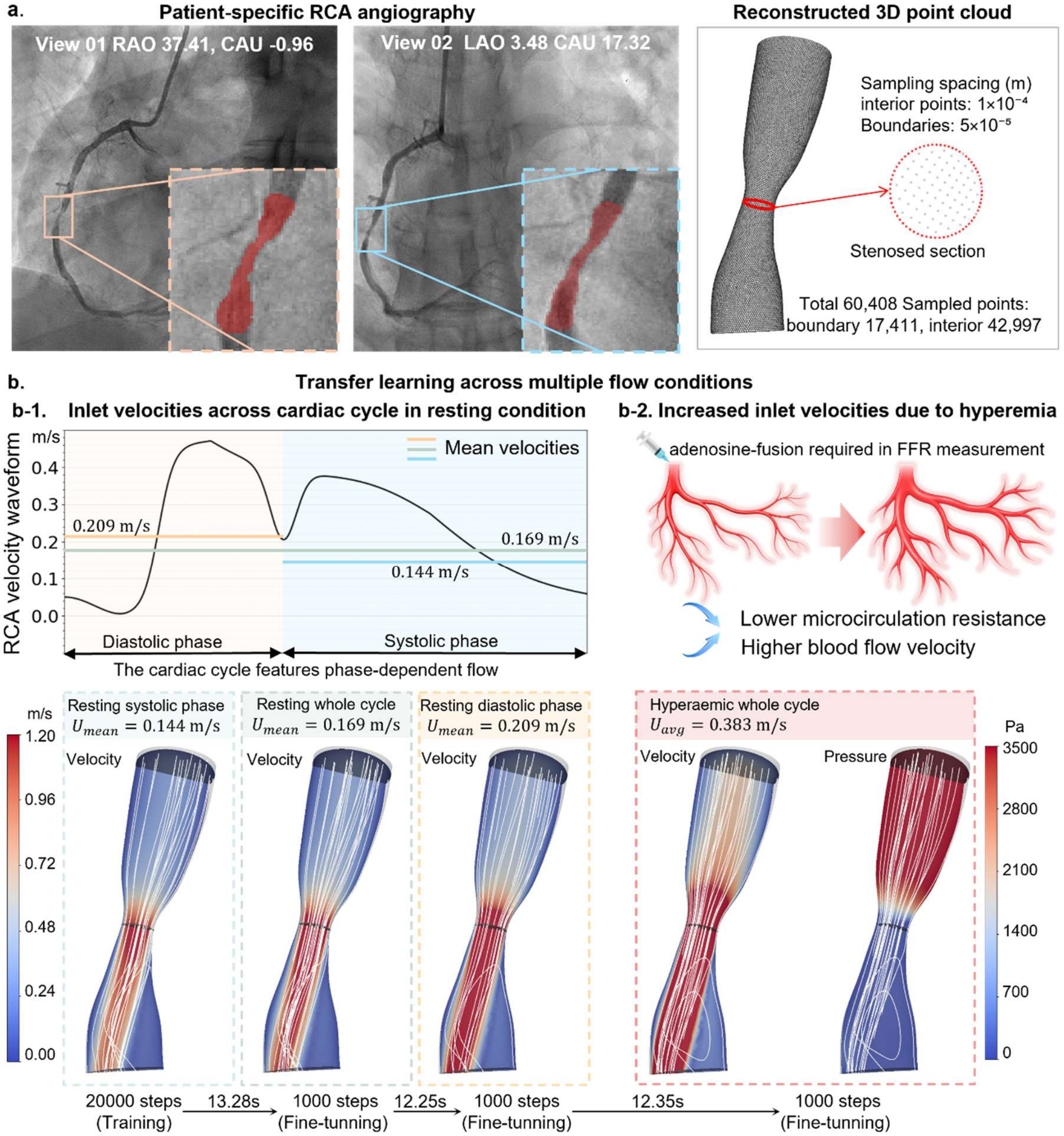

**Figure 4. Patient-specific coronary geometry reconstruction and transfer learning across multiple flow conditions.** (a) Patient-specific RCA angiography and reconstructed 3D point cloud. Dual-view right coronary artery (RCA) angiography is used to reconstruct the patient-specific three-dimensional vessel geometry, represented as a centerline-based point cloud for subsequent hemodynamic modeling. (b) Transfer learning across multiple flow conditions. Hemodynamic predictions are performed under multiple physiological flow states, including (b-1) resting systolic phase, resting whole-cycle, resting diastolic phase, and (b-2) adenosine-induced hyperemia, during which microvascular resistance decreases, resulting in accelerated coronary flow. The model is trained under the lowest inlet velocity and efficiently transferred to higher-flow conditions, enabling rapid flow prediction across diverse physiological states.

## 4.2 Consistency with Clinical FFR Diagnosis

Fractional flow reserve (FFR) is a quantitative index used to assess the functional impact of coronary artery stenosis on myocardial perfusion. Defined as the ratio of distal coronary pressure to proximal aortic pressure (FFR = $P_d/P_a$), it directly reflects the hemodynamic significance of a lesion and is widely regarded as the clinical gold standard for identifying ischemia-inducing stenosis requiring intervention [14, 16-18, 72, 73].

In clinical practice, after lesion identification by coronary angiography, a pressure wire is advanced distal to the stenosis, and maximal hyperemia is induced with adenosine (Figure 4b-2). Distal and proximal pressures are recorded to compute FFR, where FFR ≤ 0.80 indicates a hemodynamically significant lesion requiring revascularization. Pressure-wire pullback under sustained hyperemia further enables spatial localization of pressure drops, providing guidance for interventional planning. However, pressure wire–based FFR measurement remains invasive and requires pharmacologically induced hyperemia.

To reduce procedural invasiveness, we estimate FFR directly from the predicted patient-specific coronary flow fields. As illustrated in Figure 5a, the coronary artery is divided into stenotic and healthy segments based on angiographic images, and the pressure drop across the stenosis is directly obtained from the model-predicted flow field and defined as

$$\Delta P_{\mathrm{stenosis}} = P_{\mathrm{in}} - P_{\mathrm{out}}, \tag{36}$$

where $P_{\mathrm{in}}$ and $P_{\mathrm{out}}$ denote the inlet and outlet pressures of the stenotic segment.

The frictional pressure loss along the healthy vessel is analytically computed using the classical Poiseuille model under the assumption of laminar flow in a circular tube,

$$\Delta P_{\mathrm{health}} = \frac{64}{\mathrm{Re}} \cdot \frac{l}{d} \cdot \frac{\rho v^2}{2}, \tag{37}$$

where $l$ is the length of the healthy segment, $d$ the vessel diameter, $v$ the mean velocity, $\rho$ the blood density, and $Re$ the Reynolds number.

The analytically modeled healthy-vessel losses comprise the proximal segment upstream of the reconstructed lesion domain and an unmodeled distal segment extending from the reconstruction outlet to the pressure-wire sensor. The proximal segment length was measured from angiography. For distal segment, the pressure-wire sensor is positioned downstream of the stenosis, leaving a finite vessel segment between the reconstructed outlet and the pressure-reading location. Because calibrated three-dimensional sensor coordinates were unavailable, we prespecified a 15-mm outlet-to-sensor segment to account for this unmodeled distal viscous loss, consistent with the 10–20-mm post-stenotic evaluation range reported for lesion-specific image-derived FFR by Nørgaard et al. [74].

The total pressure loss used for FFR was therefore defined as

$$\Delta P = \Delta P_{\mathrm{health,\,proximal}} + \Delta P_{\mathrm{stenosis}} + \Delta P_{\mathrm{health,\,distal}}. \tag{38}$$

Finally, cuff-derived MAP was used to estimate patient-specific hyperemic aortic pressure. To account for the pressure reduction during adenosine-induced hyperemia, 6 mmHg was subtracted following the mean decrease reported by Wilson et al. [75]:

$$P_{a,i}^{\mathrm{MAP}} = \left(\frac{\mathrm{SBP}_i + 2\mathrm{DBP}_i}{3} - 6\ \mathrm{mmHg}\right) \times 133.322. \tag{39}$$

where $\mathrm{SBP}_i$, $\mathrm{DBP}_i$ denote the cuff-measured systolic and diastolic blood pressures of patient i. Using this patient-specific hyperemic aortic pressure $P_{a,i}^{\mathrm{MAP}}$, angiography-derived FFR was calculated as

$$\mathrm{FFR}_{\mathrm{MAP},i} = 1 - \frac{\Delta P}{P_{a,i}^{\mathrm{MAP}}}. \tag{40}$$

For validation, Figure 5b presents the clinically measured FFR pullback curve for this patient. After adenosine infusion, pressure-wire pullback typically shows two characteristic patterns: a nonlinear pressure drop across the stenosis caused by geometric constriction and flow acceleration, and an approximately linear decay along the proximal/distal healthy vessel dominated by viscous friction. Our model-predicted pressure profile agrees well with the clinical measurements.

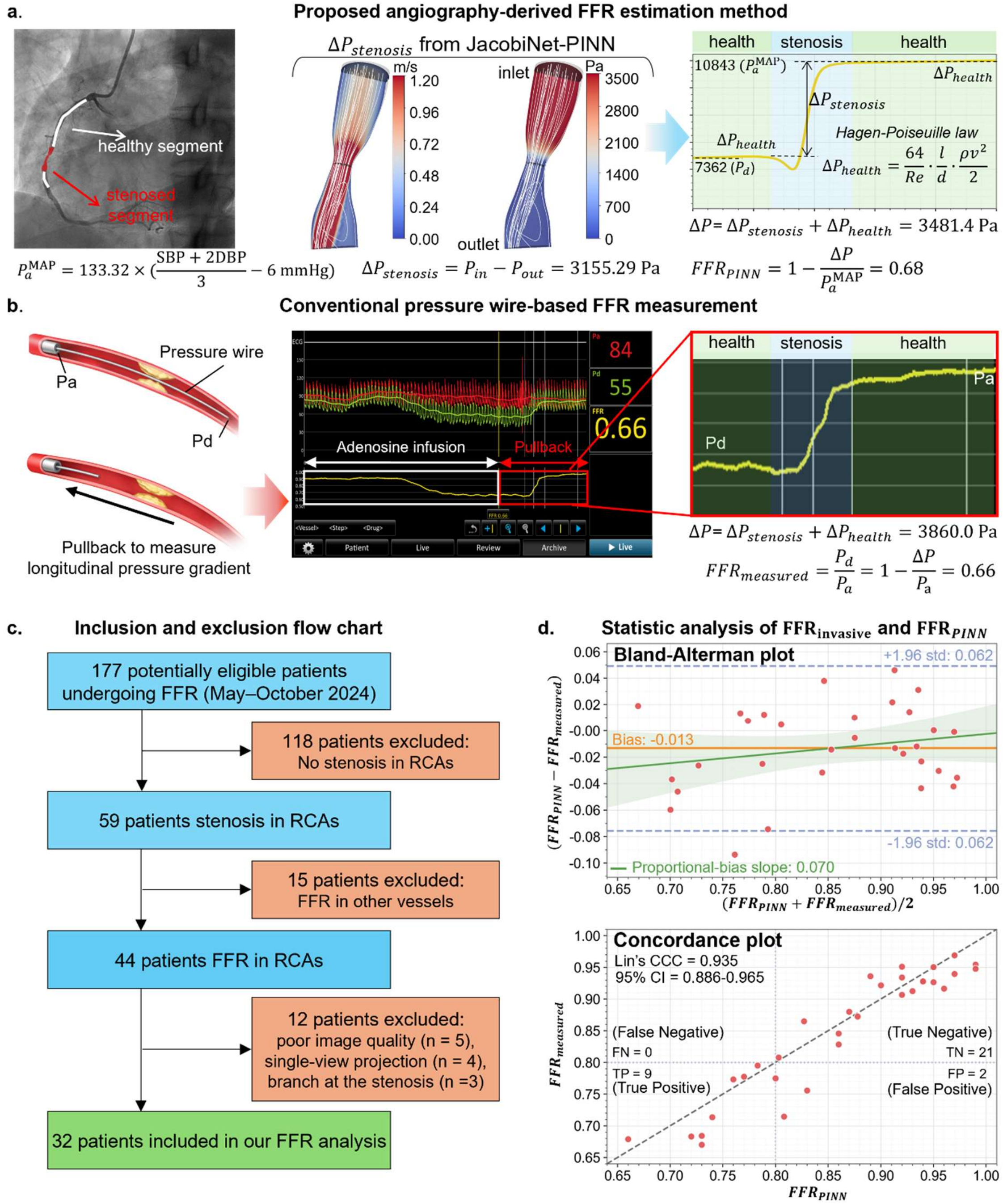

**Figure 5. Consistency between the proposed angiography-derived FFR method and clinical invasive FFR diagnosis.** (a) The proposed angiography-derived FFR computation strategy decomposes the total pressure drop into two components: the localized pressure loss across the stenotic segment, directly estimated from the model-predicted flow field, and the frictional pressure loss along the proximal/distal healthy segment, calculated using the Hagen–Poiseuille law. (b) Conventional wire-based FFR measurement using a pressure wire advanced distal to the lesion and gradually pulled back to obtain the pressure pullback curve. The measured FFR pullback curve exhibits a distinct three-stage pattern, comprising a localized pressure drop across the stenosis and a gradual frictional pressure decline along the proximal/distal healthy vessel. (c) Patient inclusion and exclusion flowchart for the clinical validation cohort. (d) Agreement between angiography-derived and invasive FFR in 32 patients. The mean difference, defined as JacobiNet–PINN FFR minus invasive FFR, was −0.013, with 95% limits of agreement from −0.075 to 0.049. Lin's CCC was 0.935 (95% CI, 0.886–0.965). The proportional-bias slope was 0.070 (95% CI, −0.053–0.193; p = 0.256).

Across 32 pairs, the positivity threshold was prespecified as FFR $\leq$ 0.80. the method achieved a diagnostic accuracy of 93.8%. Lin's concordance correlation coefficient was 0.935 (patient-level bootstrap 95% CI, 0.886–0.965; 10,000 resamples). Bland–Altman analysis shows a bias of $-0.013$, with limits of agreement between $-0.075$ to 0.049, and proportional-bias slope was 0.070 (95% CI, $-0.053$ to 0.193; $p = 0.256$). With invasive FFR $\leq 0.80$, 9 patients were positive and 23 negative; TP/FN/TN/FP were 9/0/21/2. Sensitivity was 100.0% (66.4%–100.0%), specificity 91.3% (72.0%–98.9%), PPV 81.8% (48.2%–97.7%), NPV 100.0% (83.9%–100.0%), accuracy 93.8% (30/32; 79.2%–99.2%), balanced accuracy 95.7%, and AUC 0.961 (0.884–1.000).

Figure 6 summarizes multi-condition flow predictions and FFR pressure-drop curves for 8 patients with invasive FFR values ranging from 0.70 to 0.90, including lesions near the clinical threshold of 0.80—a diagnostically challenging gray zone [14, 72]. Dual-view coronary angiography is used to reconstruct 3D vessel geometry, followed by physics-informed flow prediction under resting diastolic, resting whole-cycle, resting systolic, and adenosine-induced hyperemic conditions. The rightmost panels present angiography-derived FFR values derived from the predicted pressure fields alongside invasive measurements, showing strong agreement across all cases. **Appendices D and E** report cohort-level field-error and diagnostic results for all 32 patients.

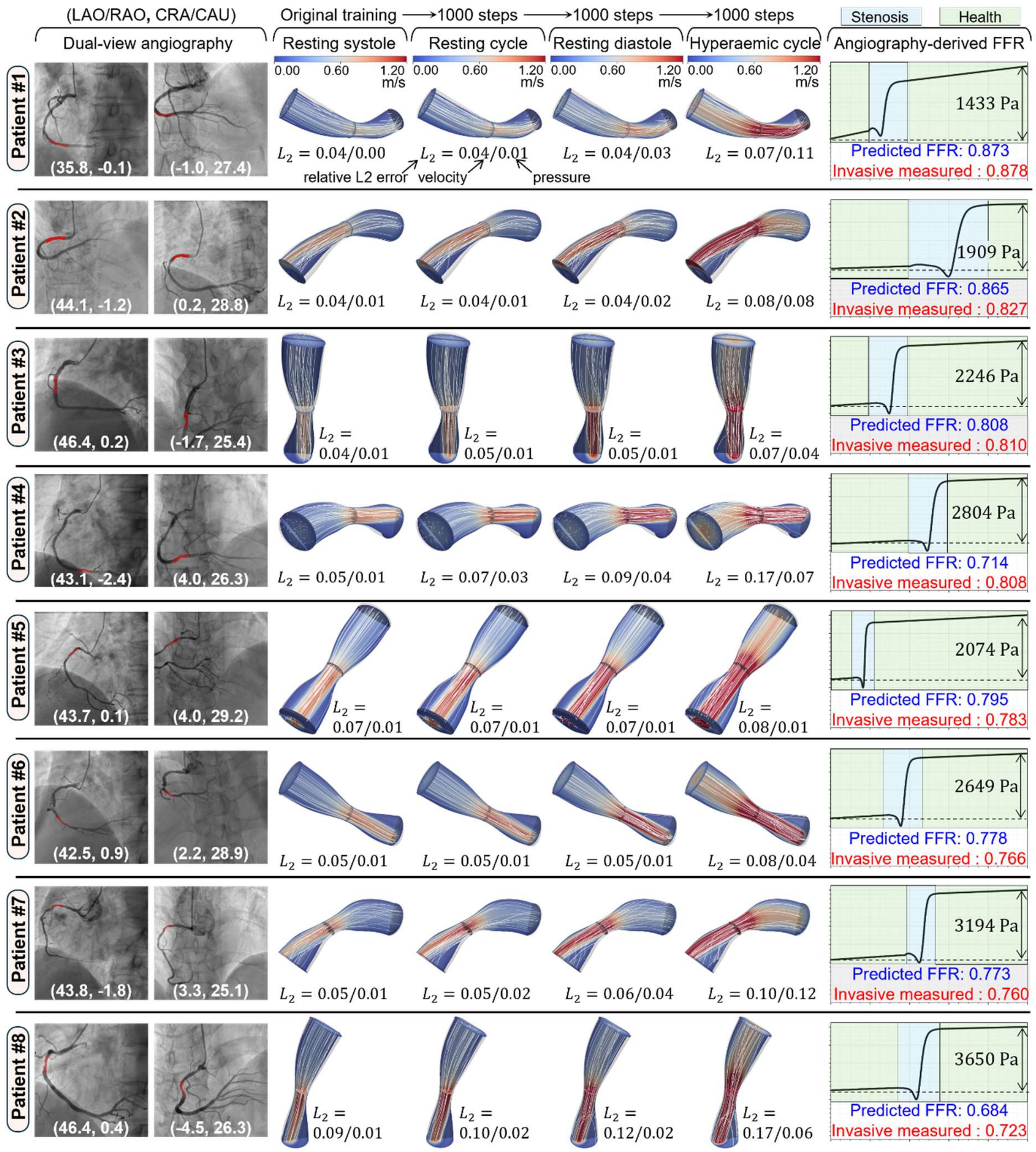

**Figure 6. Multi-condition patient-specific coronary flow reconstruction and angiography-derived FFR validation.** Representative cases from eight patients are presented, with invasive FFR values ranging from 0.7 to 0.9, including lesions around the clinical threshold of 0.80—a diagnostically challenging gray zone for functional assessment. For each case, dual-view coronary angiography is first used for three-dimensional geometric reconstruction, followed by physics-informed flow prediction under resting diastolic phase, resting whole-cycle, resting systolic phase, and adenosine-induced hyperemic conditions. The rightmost panels show the corresponding angiography-derived FFR values computed from the predicted pressure fields, alongside invasive pressure-wire measurements. Across all cases, the predicted FFR values exhibit reasonable agreement with invasive measurements, highlighting the accuracy and robustness of the proposed framework across multiple physiological flow conditions.

### 4.3 Illustrative Virtual-Revascularization Experiment

Beyond restoring coronary perfusion, stent implantation plays a critical role in reshaping the local hemodynamic environment. Wall shear stress (WSS), which is proportional to blood viscosity and the radial velocity gradient, is a key regulator of endothelial function and gene expression [76-80]. Low WSS (<4 dyne/cm²) has been reported to be associated with atherosclerotic lesion development and pro-inflammatory endothelial phenotypes [81-83], whereas excessively high WSS (>40 dyne/cm²) may induce endothelial injury and promote unstable plaque formation, thereby increasing the risk of adverse cardiovascular events [84, 85]. These associations motivate the evaluation of WSS distributions before and after stenting.

As shown in Figure 7a, in clinical practice, post-stenting angiography is typically performed using a single view to reduce radiation exposure and contrast usage, which is generally considered to be sufficient as the stented segment can be approximated as quasi-cylindrical with a nearly uniform diameter. Accordingly, geometric reconstruction in the stenting segment is formulated as a radius-updating correction process.

Based on the reconstructed three-dimensional geometry and physics-informed flow modeling, we conducted a proof-of-concept virtual-revascularization analysis of different candidate stent sizes. In Figure 7b, three stent diameters ($d$ = 2.00, 2.50, 3.00 mm) were compared based on their predicted WSS distributions. Among the tested configurations, the 2.50 mm stent yielded the most favorable outcome, with the smallest abnormal WSS area (0.07%). Post-intervention hemodynamic assessment was also explored using single-view coronary angiography (Figure 7a, middle). The projected stent length ($l = 8$ mm) and diameter ($d = 2.5$ mm) are measured from angiography. An approximate post-intervention geometry is reconstructed by locally updating the vessel radius in the stented region while preserving the centerline coordinates of the original representation $[x_m, y_m, z_m, r]$, and the corresponding flow field is then computed. As illustrated in Figure 7c, the resulting streamlines indicate a smoother and more uniform flow pattern within the stented segment.

Please note that these analyses demonstrate technical feasibility and should not be interpreted as clinical recommendations. Further translation will require explicit stent-strut modeling, validation against postoperative measurements, comparison with operator-selected devices, and association with clinical outcomes.

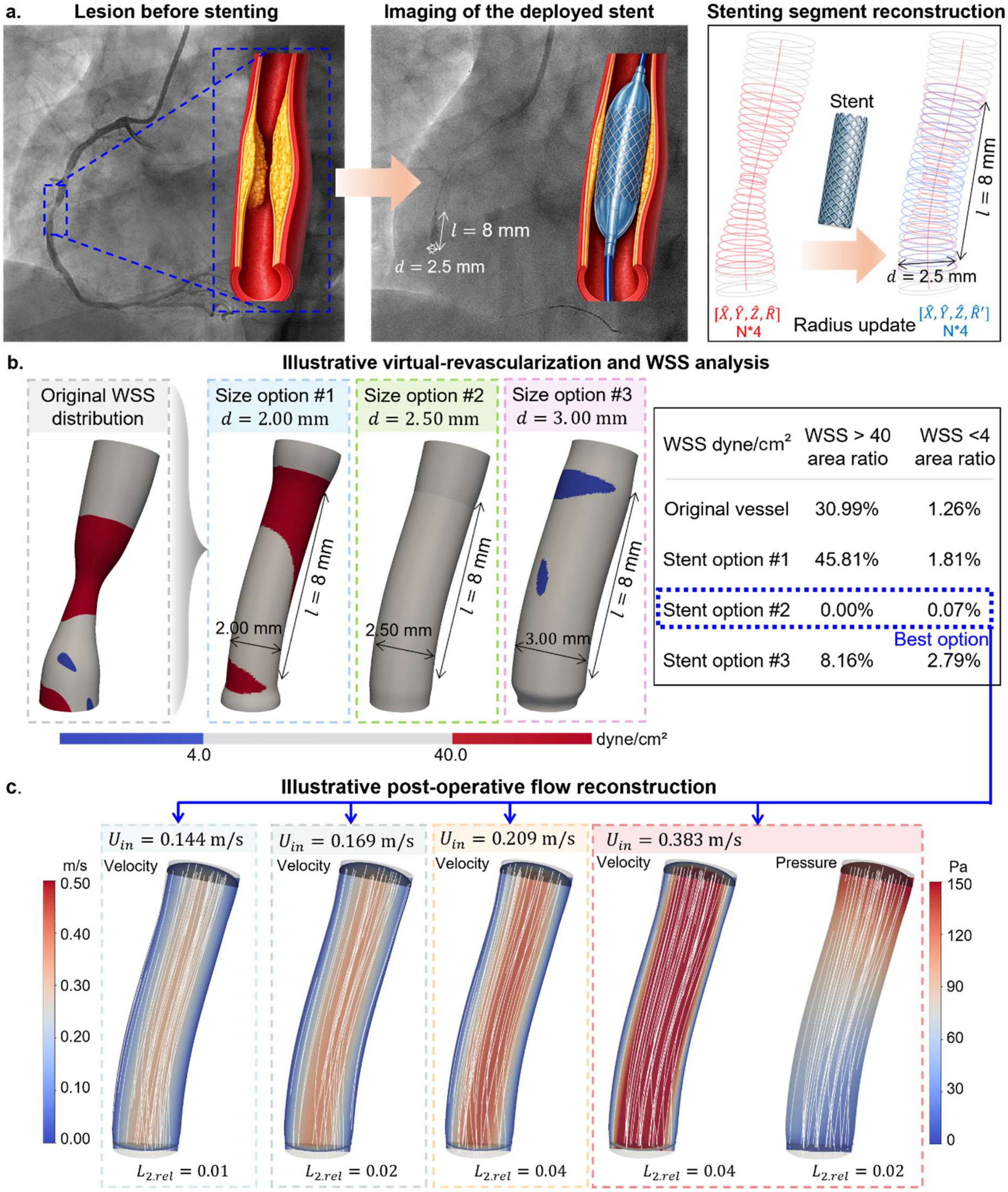

**Figure 7. Proof-of-concept virtual stenting and post-intervention hemodynamic assessment using the proposed framework.** (a) Identification and approximate reconstruction of the stented segment. Stent dimensions are extracted from single-view post-intervention angiography and used to locally update the vessel radius while preserving the original centerline. (b) Predicted WSS distributions for the original vessel and different virtual stent sizes. The intermediate configuration produced the smallest abnormal-WSS area in this controlled comparison. (c) Reconstructed post-intervention velocity and pressure fields with errors evaluated against reference CFD solutions. The predicted streamlines show smoother and more uniform flow within the stented segment under different inlet conditions.

## 5. Sparse-Data Assimilation for Flow Reconstruction

In clinical practice, boundary conditions for coronary blood flow are often incomplete or uncertain. Conventional CFD solvers require prescribed boundary conditions to obtain a well-posed problem and therefore typically rely on simplified assumptions, which lack direct physiological justification [86, 87]. Meanwhile, sparse clinical measurements—such as velocity data from Doppler guidewires or 4D-flow MRI—are typically obtained at interior vessel locations rather than at computational boundaries. Consequently, conventional CFD frameworks cannot naturally incorporate such information, leading to potential discrepancies between simulated and in vivo hemodynamics.

By contrast, the proposed learning-based framework naturally supports data fusion, enabling hemodynamic modeling without reliance on prescribed inlet flowrate. In this section, we included four synthetic cases (Figure 3) to conduct the sparse-data assimilation experiment. Vessel geometry, fluid properties, the no-slip wall condition, and the zero-gauge-pressure outlet condition were retained, with only the inlet velocity treated as unknown. The corresponding CFD solutions were sampled to emulate sparse velocity measurements, while the complete CFD fields were excluded from both model training and checkpoint selection.

As shown in Figure 8, the two experiments were designed to emulate complementary clinical velocity measurements: pointwise axial-velocity observations from intracoronary Doppler and voxel-averaged three-component velocities from single-phase 4D-flow MRI.

For simulated Doppler, the observation operator supplied only the local axial velocity, consistent with the directional sensitivity of a single-beam intracoronary Doppler guidewire [88, 89]. The N=1 configuration used one distal observation at $s = 0.70$, whereas N=3 used observations at $s = 0.30$, $0.50$, and $0.70$. We evaluated clean observations, 5% and 10% multiplicative velocity noise, and prespecified $\pm 1$ and $\pm 2$-mm coordinate--measurement mismatch.

For simulated single-phase 4D-flow MRI, the observations comprised three-component velocities voxel-averaged at 1.2-mm resolution, consistent with three-directional velocity encoding [90]. One cross-sectional layer was placed at $s = 0.70$, and three-layer sampling used $s = 0.30$, $0.50$, and $0.70$. Clean data were evaluated for both configurations, while the one-layer configuration was additionally tested with 5% and 10% multiplicative velocity noise. No additional subvoxel positional mismatch was imposed, because voxel averaging at a spatial resolution of 1.2 mm already accounted for the finite spatial support and associated within-voxel positional variation of each measurement.

The results are summarized in Figure 8. Without sparse observations, $E_U$, full-field velocity relative-$L_2$, and pressure relative-$L_2$ were $53.44 \pm 26.87\%$, $0.54 \pm 0.27$, and $0.40 \pm 0.16$, respectively. A single clean simulated Doppler observation reduced these errors to $1.89 \pm 1.71\%$, $0.05 \pm 0.03$, and $0.02 \pm 0.01$. Three clean observations did not further reduce the cohort-level means; however, their redundancy improved robustness under stronger uncertainty. At 10% noise, velocity relative-$L_2$ was $0.08 \pm 0.02$ for $N = 3$, compared with $0.13 \pm$ 0.02 for $N = 1$. Under $\pm 2$-mm positional mismatch, the corresponding values were $0.06 \pm 0.04$ and $0.14 \pm 0.10$. Simulated 4D-flow MRI observations likewise improved recovery relative to the no-data control. One clean layer yielded $E_U = 3.59 \pm 2.61\%$, velocity relative-$L_2 = 0.05 \pm 0.03$, and pressure relative-$L_2 = 0.04 \pm 0.03$. For the one-layer configuration, increasing velocity noise from 0% to 5% and 10% increased velocity relative-$L_2$ only gradually, from $0.05 \pm 0.03$ to $0.06 \pm 0.04$ and $0.06 \pm 0.04$, respectively.

These findings demonstrate the robust sparse-data assimilation capability of the proposed framework: sparse internal velocity observations enabled inference of the unknown inlet-flow scale and substantially improved full-field velocity and pressure reconstruction, with these improvements retained across the tested noise levels and positional mismatches. Although multiple noise levels and spatial misalignments were introduced to emulate the measurement noise and localization uncertainty encountered in clinical acquisitions, the simulated observations and evaluation references were derived from the same CFD solutions. Accordingly, this analysis establishes a controlled in-silico CFD-twin proof of concept; clinical validation will require independently acquired Doppler or 4D-flow MRI measurements that capture real-world acquisition and registration variability.

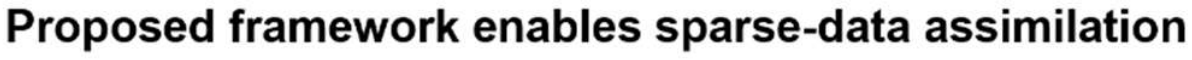

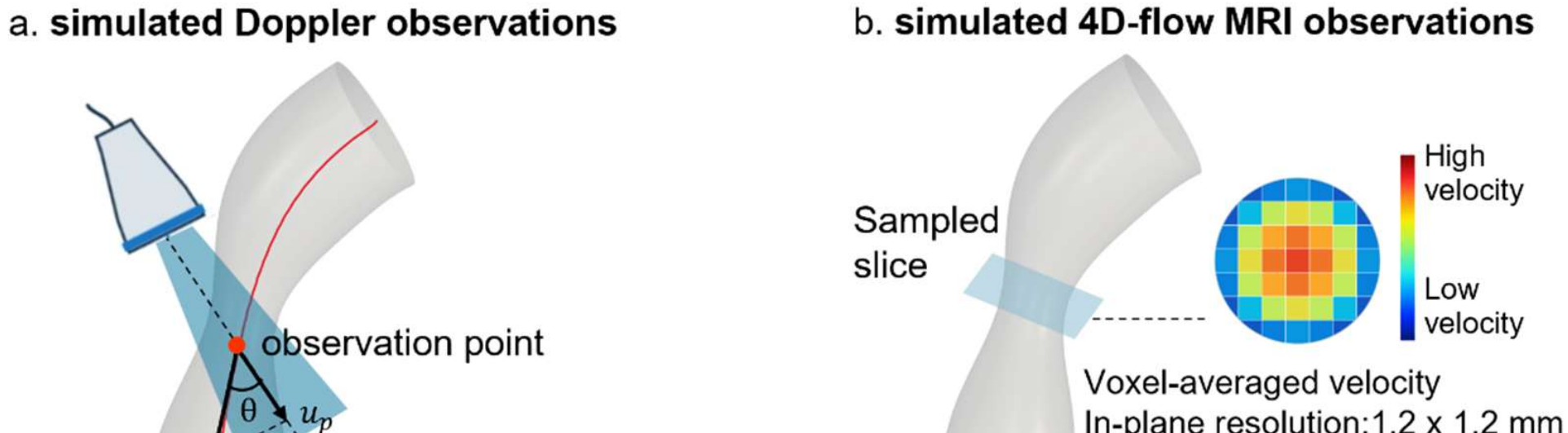

c. **Performance of data assimilation using simulated Doppler & 4D-Flow MRI observations**

| Simulated observation | Observation condition | $N$ | $E_U$, % | Velocity relative-$L_2$ | Pressure relative-$L_2$ |
|---|---|---|---|---|---|
| N/A | No sparse data | 0 | 53.44 ± 26.87 | 0.54 ± 0.27 | 0.40 ± 0.16 |
| **Doppler** | Clean | 1 | 1.89 ± 1.71 | 0.05 ± 0.03 | 0.02 ± 0.01 |
| | Clean | 3 | 3.40 ± 3.82 | 0.05 ± 0.04 | 0.03 ± 0.03 |
| | 5% velocity noise | 1 | 7.18 ± 1.25 | 0.08 ± 0.02 | 0.07 ± 0.01 |
| | 5% velocity noise | 3 | 5.70 ± 2.10 | 0.07 ± 0.03 | 0.05 ± 0.02 |
| | 10% velocity noise | 1 | 12.56 ± 1.01 | 0.13 ± 0.02 | 0.12 ± 0.02 |
| | 10% velocity noise | 3 | 7.97 ± 0.71 | 0.08 ± 0.02 | 0.07 ± 0.01 |
| | ±1-mm position mismatch | 1 | 5.28 ± 3.09 | 0.06 ± 0.03 | 0.05 ± 0.03 |
| | ±1-mm position mismatch | 3 | 2.54 ± 4.30 | 0.05 ± 0.04 | 0.03 ± 0.04 |
| | ±2-mm position mismatch | 1 | 13.00 ± 10.67 | 0.14 ± 0.10 | 0.12 ± 0.11 |
| | ±2-mm position mismatch | 3 | 4.40 ± 3.36 | 0.06 ± 0.04 | 0.04 ± 0.04 |
| **4D-flow MRI** | Clean, 1 layer | 1 | 3.59 ± 2.61 | 0.05 ± 0.03 | 0.04 ± 0.03 |
| | Clean, three layers | 3 | 4.97 ± 5.15 | 0.06 ± 0.05 | 0.05 ± 0.05 |
| | 5% velocity noise | 1 | 3.66 ± 3.38 | 0.06 ± 0.04 | 0.04 ± 0.03 |
| | 10% velocity noise | 1 | 4.03 ± 4.07 | 0.06 ± 0.04 | 0.04 ± 0.04 |

$E_U = |\hat{U} - U_{\mathrm{ref}}|/U_{\mathrm{ref}}$ denotes the relative error in the inferred mean inlet velocity.
For simulated Doppler measurements, $N$ denotes the number of point observations; for simulated 4D-flow MRI, $N$ denotes the number of cross-sectional layers.

**Figure 8. Sparse-data assimilation using simulated Doppler and 4D-flow MRI observations.** (a) Schematics of angle-corrected Doppler point measurements; (b) Schematics of voxel-averaged 4D-flow MRI measurements; (c) Reconstruction performance under different observation numbers, noise levels, and positional mismatches. Values are mean ± standard deviation across four synthetic cases. N denotes Doppler points or MRI layers, and *EU* denotes the relative error in the inferred inlet velocity.

# 6. Discussion & conclusion

The results highlight several key advantages. First, the geometric embedding via JacobiNet alleviates the stiffness commonly encountered in unsupervised PINNs for complex vascular domains, leading to fast convergence and well stability even under high Reynolds number conditions. Second, the hard-constrained trial functions eliminate the need for boundary penalties, ensuring strict satisfaction of no-slip, inlet, and outlet conditions. By embedding physical priors into the network design, the framework also achieves efficient parameter transfer across multiple physiological flow states. Third, the framework naturally incorporates sparse interior clinical measurements, such as intracoronary Doppler flow velocity or MRI, and remains well-posed even when inlet velocity scales are unknown. Collectively, these features allow the proposed approach to overcome fundamental limitations of conventional CFD pipelines and existing learning-based surrogates, offering a flexible and clinically aligned alternative for coronary functional assessment.

The full pipeline—including multi-condition flow reconstruction, FFR prediction, illustrative virtual-revascularization comparison—completes within 20 minutes per case on average. Further algorithmic optimization and hardware acceleration could potentially reduce runtime. For example, transitioning to a JAX-based implementation for more efficient higher-order automatic differentiation [91], incorporating advanced operator-learning frameworks such as DeepONet [92] or Fourier neural operator [93], and deploying industrial-grade GPUs rather than development-level consumer hardware (e.g., RTX 4090D) may further enhance computational efficiency and scalability.

Moreover, reconstruction was evaluated within prespecified acquisition ranges using two complementary RCA views. To enhance robustness across imaging protocols and projection angles, incorporating acquisition parameters into the network will be an important direction for future research. DICOM-based pixel-spacing/SID/SOD calibration harmonized the reference-plane scale, although residual effects associated with vessel depth, foreshortening, cardiac motion, and out-of-plane anatomy warrant further investigation. The centerline-plus-scalar-radius representation accommodated axis-offset eccentric stenoses (Appendix-F) while maintaining locally circular cross-sections. Future work will incorporate more flexible lumen representations and extend the framework from unbranched, single-lesion RCA segments to branched RCA and LAD/LCx anatomies, as well as tandem, diffuse, and calcified lesions, through branch-aware reconstruction, patient-specific flow allocation, and multi-outlet boundary conditions.

Furthermore, despite the high overall diagnostic accuracy, the confidence intervals for sensitivity and predictive values remained relatively wide because of the modest size of the retrospective clinical cohort. Larger external cohorts are therefore essential for translating the proposed framework into routine clinical decision support. Moreover, the virtual-revascularization and sparse-data experiments were designed as controlled proofs of concept. CFD agreement supports the computational feasibility of WSS-based virtual device comparison, with clinical outcome validation reserved for future studies. Besides, the current framework assumes steady-state, Newtonian flow. Extending the model to fully transient simulations and non-Newtonian rheology may enhance physiological fidelity, particularly in complex or severely diseased conditions. Addressing these challenges will be critical for advancing physics-informed coronary modeling toward robust and routine clinical deployment.

## Conflict of Interest

The authors declare that the research was conducted in the absence of any commercial or financial relationships that could be construed as a potential conflict of interest. No patent applications relate to the pipeline described in this study or to the GitHub repository.

## Data availability

The source code and synthetic data generator are currently being organized for public release and will be made available at https://github.com/xchenim/JacobiNet_bloodflow on September 15, 2026. Released materials include the reconstruction and JacobiNet–PINN pipelines, frozen model weights, and synthetic dataset.

## Acknowledgements

This work was supported by the HKUST Startup Funding.

## Contributions

X.C. and W.H. conceived and designed the study. X.C. developed the algorithms, implemented the computational framework, and drafted the manuscript. H. L., Q.L. and M.C. acquired the clinical data. X.C., J.Y., G.H. and Q.Y. performed the experiments and analyzed the data. All authors discussed the results, contributed to revising the manuscript, and approved the final version.

# Appendix-A. Synthetic Dataset Generation

Clinical coronary angiography is limited in scale, and cardiac-motion-induced deformation can introduce temporal misalignment between views. We therefore constructed a controlled synthetic right coronary artery (RCA) dataset for training and evaluating the dual-view three-dimensional reconstruction network. The dataset contains 50,000 independently generated RCA geometries and their paired angiographic projections.

### A.1 Clinical source and ethical approval

The generation framework was adapted from the coronary-geometry synthesis procedure described previously [1]. De-identified coronary computed-tomography angiography data from 10 patients acquired at the University of Michigan Hospital were retrospectively collected under approval from the University of Michigan Institutional Review Board for Medicine (IRBMED; HUM00155491). Three-dimensional coronary centerlines were delineated with CRIMSON. Vessel length, branch location, lumen dimension, and radius ranges extracted from these data were used to define physiologically plausible sampling intervals; the clinical geometries themselves were not used as reconstruction-network training targets.

### A.2 Geometry generation

New centerlines were generated by uniformly sampling control points within the clinically derived parameter ranges, applying cubic B-spline interpolation, and resampling each curve to 140 equally spaced arc-length locations. Main-vessel length was sampled uniformly from 130 to 150 mm [2]. The radius profile followed a longitudinal linear taper with small random perturbations.

One stenosis was inserted into each synthetic vessel. Diameter reduction ranged from 5% to 90%, covering mild through severe disease [3,4]. The lesion center was sampled uniformly along the normalized centerline coordinate, $s \in [0.2, 0.8]$, to avoid the inlet and outlet. Lesion length was sampled in physical space from 10.7 to 20.7 mm [5–7] and converted to the corresponding arc-length interval on each centerline. Each generated geometry was represented by its paired three-dimensional centerline coordinates and local radii. Random rigid rotations, shear transformations, and low-frequency non-rigid deformations were subsequently applied to vary orientation, curvature, and torsion while preserving vessel topology.

### A.3 Dual-view projection generation

A cone-beam model generated two projections of each synthetic geometry on a $512 \times 512$ canvas. The angles below follow the X-ray-source-side convention used by the synthetic generator; detector-side DICOM angles describe the opposing direction.

| Projection | Prespecified source-side configuration | Sampling rule |
|---|---|---|
| View 1 | RAO 40° with secondary CRA/CAU 0° | RAO 40° $\pm$ 10°; CRA/CAU 0° $\pm$ 5° |
| View 2 | RAO/LAO 0° with CAU 25° | RAO/LAO 0° $\pm$ 5°; CAU 25° $\pm$ 10° |

All angular components were drawn independently from bounded uniform distributions within the stated intervals. These configurations were selected to approximate commonly used complementary RCA projections [8,9]. The imaging geometry was fixed at a source-to-image distance of 1,100 mm, a source-to-object distance of 765 mm, and an imager pixel spacing of 0.258390625 mm/pixel.

### A.4 Image preprocessing

The $512 \times 512$ images are projection-generation canvases rather than network-input dimensions. A Euclidean distance transform was applied to each binary vessel projection to enhance the continuous vascular structure and suppress the background. Each transformed view was then converted to grayscale, centrally zero-padded only when either dimension was smaller than 128 pixels, centrally cropped to exactly $128 \times 128$, and normalized to $[0, 1]$ by division by 255. No interpolation-based resizing was performed. The two single-channel views were retained separately and supplied to the shared-weight encoder as a tensor of shape $2 \times 1 \times 128 \times 128$.

### A.5 Dataset partition

The 50,000 samples were divided into 40,000 training, 5,000 validation, and 5,000 test cases. The validation set was used for model and checkpoint selection. The test set remained independent and was used only for final evaluation after the reconstruction configuration had been fixed. Each sample paired the two preprocessed projections with the corresponding three-dimensional centerline and radius targets.

# Appendix-B. Experimental Configurations

## B.1 Attention-enhanced network for three-dimensional coronary reconstruction

Each sample comprised two separately retained $1 \times 128 \times 128$ grayscale distance-transform images. A shared one-channel input adapter and an ImageNet-V1-pretrained ResNet-50 backbone [10,11] processed the views with identical weights. For each input, the encoder returned `layer2`, `layer3`, and `layer4` feature maps; the `layer4` output had shape $2048 \times 4 \times 4$.

A convolutional block attention module (CBAM) [12] was applied to `layer4`,

$$\tilde{\mathbf{F}} = \mathbf{M}_s \otimes (\mathbf{M}_c \otimes \mathbf{F}),$$

where channel and spatial attention were applied sequentially, with a channel-reduction ratio of 8 and a spatial kernel size of 7. CBAM preserved the $2048 \times 4 \times 4$ dimensions. Global average pooling generated a 2,048-dimensional vector per view, and the two vectors were concatenated into a 4,096-dimensional representation.

The centerline head comprised three 512-unit Linear--LayerNorm--GELU blocks followed by a linear layer and predicted 12 ordered three-dimensional points. The radius branch combined a global three-block, 256-unit MLP with features from `layer2`, `layer3`, and `layer4`. Each scale was projected to 64 channels, spatial maximum pooling was applied separately to both views, and the concatenated context was passed through a three-block, 256-unit residual MLP to refine the global radius estimate. CBAM feature recalibration, multiscale pooling, and residual radius refinement together constitute the integral MS-CBAM design used in the model.

The first radius latent variable represented standardized log inlet radius, and the remaining variables represented log-radius ratios relative to the inlet. Exponentiation yielded 12 positive radii paired pointwise with the 12 centerline locations. The training targets were obtained from the original 140-point geometry using a fixed 12-point rule that retained the global minimum-radius location and applied the same indices to coordinates and radii.

The batch size was 128. AdamW used weight decay $10^{-4}$. Initial learning rates were $3 \times 10^{-4}$ for the task heads, $3 \times 10^{-5}$ for CBAM, and $10^{-5}$ for the earlier backbone layers; Stage 2 exposed only the radius head at $3 \times 10^{-4}$. A 5% linear warm-up was followed by cosine decay to $10^{-6}$, and the global gradient norm was clipped at 1.0. Checkpoint selection used only the 5,000-case validation set; the 5,000-case test set was used only after configuration selection. The final model contained approximately $2.90 \times 10^{7}$ parameters.

## B.2 JacobiNet--PINN for three-dimensional coronary flow

The geometry was rigidly aligned so that the inlet plane was normal to the $z$-axis and inflow was directed along positive $z$. JacobiNet had architecture $3 \rightarrow 128 \rightarrow 128 \rightarrow 2$, with SiLU activations in its two hidden layers and 17,282 parameters. Adam used seed 99 and cosine learning-rate decay from $10^{-3}$ to $10^{-5}$, with a cap of 100,000 steps. JacobiNet was frozen during PINN training.

Physical and geometry-parameterized coordinates were encoded separately with fixed random Fourier features (RFFs) [13,14]. The RFF used 32 Gaussian directions for Cartesian coordinates and 32 independent directions for reference coordinates; each branch therefore yielded 64 sine/cosine features. The base Gaussian scale was $\sigma = 10$, with frequency factors 1 for physical coordinates and 4 for geometry coordinates. The two encodings were concatenated into a shared 128-dimensional feature vector. Velocity and pressure were then predicted by separate subnetworks: four 128-unit SiLU hidden layers with three outputs for velocity, and three 128-unit SiLU hidden layers with one output for pressure. JacobiNet and the two field networks contained approximately $1.33 \times 10^5$ parameters before JacobiNet was frozen.

Interior and boundary point sets were generated once for each geometry and remained fixed. A seeded sampler traversed randomized permutations of the complete interior pool in batches of 8,192, without replacement or omission within each pass, and reshuffled only after exhaustion. Because boundary conditions were encoded analytically, boundary points were excluded from training and used only for evaluation. A separate fixed set of 2,048 interior points was used for label-free validation and checkpoint selection; CFD labels were never used for either purpose.

Across the 32 patient-specific geometries, fixed interior PDE pools contained 20,849--161,710 points (mean 57,118), and boundary sets contained 10,646--41,483 points (mean 19,694). Case-wise counts are given in Appendix-D. PINN optimization followed [15,16] and used Adam, seed 99, IEEE FP32 arithmetic with TF32 disabled, and cosine learning-rate decay from $10^{-3}$ to $10^{-5}$. There was no early stopping. Fixed horizons were assigned by stenosis severity:

| **Diameter-stenosis stratum of processed geometry** | **Optimizer steps** |
| --- | --- |
| 0--<25% | 10,000 |
| 25--<50% | 20,000 |
| 50--<70% | 30,000 |
| 70--<100% | 40,000 |

### B.3 CFD reference configuration

Reference fields were computed with ANSYS Fluent 2024 R1 (v24.1) in three-dimensional double precision. Blood was modeled as a steady, incompressible Newtonian fluid governed by the laminar Navier--Stokes equations, with $\rho = 1{,}060\ \mathrm{kg/m^3}$ and $\mu = 0.004$ Pas. The wall was rigid and no-slip, the inlet profile was parabolic, and outlet gauge pressure was 0 Pa.

| **Component** | **Fixed CFD setting** |
| --- | --- |
| Mesh workflow | Fluent Meshing Watertight Geometry |
| Surface mesh | Curvature-based triangles; wall size 0.075 mm; growth rate 1.10; curvature normal angle 20° |

| Component | Fixed CFD setting |
|---|---|
| Core mesh | Unstructured tetrahedra; maximum size 0.600 mm; growth rate 1.20 |
| Near-wall mesh | 10 wedge-prism layers; first layer 0.010 mm; growth rate 1.10; prescribed total thickness 0.159 mm; connected tetrahedral core |
| Near-wall treatment | Laminar model; no wall function and no $y^+$ target; prism geometry audited directly |
| Coupling and discretization | SIMPLEC; least-squares cell-based gradients; second-order pressure; second-order-upwind momentum |
| Under-relaxation and initialization | Pressure 0.3; momentum 0.7; hybrid initialization |
| Solver stopping rule | Continuity residual $\leq 10^{-5}$ and each momentum residual $\leq 10^{-7}$; otherwise, maximum 2,000 iterations |
| Final solution acceptance | Continuity residual $\leq 10^{-3}$, each momentum residual $\leq 10^{-5}$, and relative mass imbalance $\leq 10^{-3}$ |
| Mesh acceptance | Fluent mesh check passed; minimum orthogonal quality $\geq 0.04$; maximum surface skewness $\leq 0.80$; inlet area, outlet area, wall area, and fluid-volume differences each $\leq 1\%$; complete 10-layer prism columns |

## B.4 Mesh-independence study and CFD audit

Four prespecified held-out geometries, one per stenosis stratum, were evaluated at core maximum sizes 1.350, 0.900, 0.600, 0.400, and 0.300 mm. Wall size was one eighth of core size and the prism specification was unchanged. Pressure drop was area-averaged inlet static gauge pressure minus area-averaged outlet static gauge pressure; outlet mass flow was absolute, and mean WSS was the wall-area average. Displayed values are rounded; GCI used full precision.

| Case | Diameter stenosis | Core size (mm) | Cells | Pressure drop (Pa) | Outlet mass flow (kg/s) | Mean WSS (Pa) |
|---|---|---|---|---|---|---|
| 00370 | 18.67% | 0.300 | 5,690,887 | 26.144 | 0.00233028 | 1.5330 |
| | | 0.400 | 3,154,777 | 26.134 | 0.00233037 | 1.5331 |
| | | 0.600 | 1,363,043 | 26.121 | 0.00233060 | 1.5335 |
| | | 0.900 | 582,355 | 26.109 | 0.00233114 | 1.5340 |
| | | 1.350 | 245,943 | 26.100 | 0.00233235 | 1.5353 |
| 01105 | 33.52% | 0.300 | 2,295,154 | 126.131 | 0.00131512 | 4.5014 |
| | | 0.400 | 1,263,399 | 126.054 | 0.00131521 | 4.5000 |
| | | 0.600 | 541,243 | 125.866 | 0.00131544 | 4.4965 |
| | | 0.900 | 229,128 | 125.631 | 0.00131597 | 4.4914 |
| | | 1.350 | 96,242 | 124.908 | 0.00131713 | 4.4787 |
| 03889 | 52.26% | 0.300 | 1,897,733 | 386.608 | 0.00107266 | 6.8567 |
| | | 0.400 | 1,043,841 | 386.365 | 0.00107275 | 6.8586 |
| | | 0.600 | 444,727 | 385.886 | 0.00107297 | 6.8603 |
| | | 0.900 | 187,831 | 384.709 | 0.00107350 | 6.8656 |
| | | 1.350 | 78,625 | 382.368 | 0.00107471 | 6.8922 |
| 04027 | 70.97% | 0.300 | 2,488,458 | 2,557.058 | 0.00188474 | 14.7921 |
| | | 0.400 | 1,372,056 | 2,554.850 | 0.00188482 | 14.8017 |
| | | 0.600 | 588,974 | 2,547.221 | 0.00188506 | 14.8364 |
| | | 0.900 | 250,687 | 2,528.046 | 0.00188558 | 14.8882 |
| | | 1.350 | 105,383 | 2,469.989 | 0.00188676 | 15.0668 |

GCI followed the generalized Celik procedure [17], with safety factor 1.25 and $h = (V/N)^{1/3}$. The 0.600-mm grid was assessed with the 0.400/0.600/0.900-mm triplet.

| Case | Stenosis stratum | Cells at 0.600 mm | Pressure-drop GCI | Outlet-mass-flow GCI | Mean-WSS GCI | Assessment |
|---|---|---|---|---|---|---|
| 00370 | 0--24% | 1,363,043 | 0.431% | 0.022% | 0.291% | Accepted |
| 01105 | 25--49% | 541,243 | 0.963% | 0.038% | 0.341% | Accepted |
| 03889 | 50--69% | 444,727 | 0.265% | 0.048% | 0.049% | Accepted |
| 04027 | 70--99% | 588,974 | 0.628% | 0.028% | 0.907% | Accepted |

All reported GCI magnitudes were below 1%. The 0.900-mm grid was not selected because pressure-drop GCI reached 1.432% in the most severe case. The 0.600-mm setting was retained.

All 100 production reference meshes passed quality checks, with $703{,}888 \pm 252{,}471$ cells per case (median 662,733; range 290,117--1,544,671), minimum orthogonal quality 0.098, and maximum surface skewness 0.61. All 100 CFD solutions met final acceptance; cohort maxima were $1.55 \times 10^{-4}$ for continuity residual, $2.164 \times 10^{-7}$ for momentum residual, and $3.986 \times 10^{-7}$ for relative mass imbalance.

## Appendix-C. Synthetic dataset ablation and benchmarking

A cohort of 100 geometries was fixed from the held-out 5,000-case reconstruction test set, with 25 cases in each label-defined stenosis stratum (0–24%, 25–49%, 50–69%, and 70–99%). The same cohort was used for all flow-solver validation, ablation, and convergence analyses. CFD results were used only for post-training evaluation and did not enter PINN training, stopping, or checkpoint selection.

### C.1 Flow-solver component ablation

All variants used the same case geometries, data split, random seed, collocation batch size of 8,192, severity-specific training horizon, FP32 precision policy, label-free checkpoint-selection rule, CFD references, and evaluation metrics. **JacobiNet–PINN (full model)** combined JacobiNet, separate RFF encodings of the natural and transformed coordinates, separate velocity and pressure networks, and analytical constraints. **w/o random Fourier features** removed RFF while retaining JacobiNet and analytical constraints; **w/o JacobiNet** used Cartesian RFF, separate velocity and pressure networks, and soft boundary losses; **w/o $(1-r^2)$ trial function** replaced the radial hard-wall envelope with a soft no-slip loss; **Unified spectral encoding** applied one joint RFF encoding to all coordinates; **Shared U/p trunk** used one joint-output network; **Soft boundary constraints** replaced the trial functions with boundary-loss terms; and **Vanilla baseline** used raw Cartesian coordinates, one shared network, and soft boundary losses without JacobiNet or RFF.

**Table C1. Component ablation of the JacobiNet–PINN flow solver. Values are case-wise mean ± sample SD.**

| Stenosis severity | Model/control | Velocity relative-$L_2$ error | Pressure relative-$L_2$ error | Pressure-drop error, Pa |
|---|---|---|---|---|
| All | JacobiNet–PINN (full model) | **0.0948 ± 0.1055** | **0.0236 ± 0.0414** | **64.16 ± 168.24** |
| | w/o random Fourier features | 0.1625 ± 0.2541 | 0.1429 ± 0.3034 | 525.90 ± 1289.41 |
| | w/o JacobiNet | 0.1210 ± 0.1413 | 0.0967 ± 0.1385 | 274.71 ± 748.50 |
| | w/o $(1-r^2)$ trial function | 0.1220 ± 0.1403 | 0.0374 ± 0.0519 | 97.86 ± 229.41 |
| | Unified spectral encoding | 0.1103 ± 0.1303 | 0.0399 ± 0.0727 | 124.95 ± 333.82 |
| | Shared U/p trunk | 0.1051 ± 0.1243 | 0.0301 ± 0.0618 | 93.35 ± 292.37 |
| | Soft boundary constraints | 0.1311 ± 0.1499 | 0.0462 ± 0.0611 | 115.41 ± 264.24 |
| | Vanilla baseline | 0.3091 ± 0.3409 | 0.3745 ± 0.3854 | 915.72 ± 1546.74 |
| 0–24% | JacobiNet–PINN (full model) | 0.0270 ± 0.0020 | **0.0075 ± 0.0023** | **0.26 ± 0.27** |
| | w/o random Fourier features | **0.0268 ± 0.0020** | 0.0098 ± 0.0033 | 0.45 ± 0.39 |
| | w/o JacobiNet | 0.0280 ± 0.0032 | 0.0284 ± 0.0202 | 1.63 ± 1.67 |
| | w/o $(1-r^2)$ trial function | 0.0269 ± 0.0020 | 0.0103 ± 0.0041 | 0.51 ± 0.44 |
| | Unified spectral encoding | 0.0270 ± 0.0020 | 0.0076 ± 0.0023 | 0.27 ± 0.30 |
| | Shared U/p trunk | 0.0270 ± 0.0020 | 0.0077 ± 0.0023 | 0.29 ± 0.30 |
| | Soft boundary constraints | 0.0269 ± 0.0020 | 0.0136 ± 0.0079 | 0.68 ± 0.71 |
| | Vanilla baseline | 0.0529 ± 0.0417 | 0.1201 ± 0.1104 | 7.92 ± 10.64 |
| 25–49% | JacobiNet–PINN (full model) | 0.0313 ± 0.0038 | 0.0071 ± 0.0021 | 0.43 ± 0.37 |
| | w/o random Fourier features | **0.0310 ± 0.0037** | 0.0081 ± 0.0028 | 0.69 ± 0.66 |
| | w/o JacobiNet | 0.0347 ± 0.0094 | 0.0307 ± 0.0271 | 5.30 ± 7.39 |

| Stenosis severity | Model/control | Velocity relative-$L_2$ error | Pressure relative-$L_2$ error | Pressure-drop error, Pa |
|---|---|---|---|---|
| | w/o $(1-r^2)$ trial function | 0.0314 ± 0.0044 | 0.0108 ± 0.0062 | 1.60 ± 2.38 |
| | Unified spectral encoding | 0.0313 ± 0.0038 | 0.0076 ± 0.0029 | 0.65 ± 1.34 |
| | Shared U/p trunk | 0.0312 ± 0.0038 | **0.0070 ± 0.0021** | **0.38 ± 0.27** |
| | Soft boundary constraints | 0.0328 ± 0.0076 | 0.0157 ± 0.0130 | 2.63 ± 4.14 |
| | Vanilla baseline | 0.0765 ± 0.0663 | 0.1188 ± 0.1160 | 21.48 ± 30.21 |
| 50–69% | JacobiNet–PINN (full model) | **0.0838 ± 0.0539** | **0.0128 ± 0.0128** | **11.68 ± 20.56** |
| | w/o random Fourier features | 0.1041 ± 0.1072 | 0.0406 ± 0.1077 | 60.97 ± 210.54 |
| | w/o JacobiNet | 0.1529 ± 0.1272 | 0.0918 ± 0.0885 | 98.94 ± 119.30 |
| | w/o $(1-r^2)$ trial function | 0.1450 ± 0.1229 | 0.0406 ± 0.0531 | 43.03 ± 64.70 |
| | Unified spectral encoding | 0.1006 ± 0.0808 | 0.0208 ± 0.0227 | 21.23 ± 34.14 |
| | Shared U/p trunk | 0.0932 ± 0.0620 | 0.0183 ± 0.0215 | 17.33 ± 25.72 |
| | Soft boundary constraints | 0.1669 ± 0.1369 | 0.0521 ± 0.0635 | 54.27 ± 76.85 |
| | Vanilla baseline | 0.3131 ± 0.2707 | 0.3439 ± 0.3496 | 414.77 ± 582.97 |
| 70–99% | JacobiNet–PINN (full model) | **0.2373 ± 0.1129** | **0.0668 ± 0.0655** | **244.27 ± 266.83** |
| | w/o random Fourier features | 0.4882 ± 0.3220 | 0.5133 ± 0.4205 | 2041.49 ± 1902.88 |
| | w/o JacobiNet | 0.2684 ± 0.1601 | 0.2359 ± 0.2015 | 992.98 ± 1254.54 |
| | w/o $(1-r^2)$ trial function | 0.2846 ± 0.1414 | 0.0881 ± 0.0638 | 346.30 ± 354.90 |
| | Unified spectral encoding | 0.2823 ± 0.1371 | 0.1236 ± 0.1069 | 477.65 ± 534.27 |
| | Shared U/p trunk | 0.2689 ± 0.1405 | 0.0872 ± 0.1034 | 355.42 ± 506.31 |
| | Soft boundary constraints | 0.2978 ± 0.1491 | 0.1033 ± 0.0750 | 404.06 ± 405.53 |
| | Vanilla baseline | 0.7937 ± 0.1747 | 0.9151 ± 0.1546 | 3218.70 ± 1431.85 |

Boldface indicates the lowest unrounded mean within each severity stratum. JacobiNet–PINN achieved the best overall result and remained the most robust in moderate-to-severe stenosis. In the 70–99% stratum, removing RFF increased pressure relative-$L_2$ from 0.0668 to 0.5133 and pressure-drop error from 244.27 to 2,041.49 Pa. Removing JacobiNet increased the same errors to 0.2359 and 992.98 Pa. Unified spectral encoding and a shared U/p trunk produced smaller but consistent deteriorations, supporting separate spectral representations and velocity–pressure networks. Soft boundary constraints and removal of the radial envelope increased pressure-drop error to 404.06 and 346.30 Pa, respectively. The vanilla baseline was substantially less accurate, reaching a pressure relative-$L_2$ error of 0.9151 and a pressure-drop error of 3,218.70 Pa in severe stenosis. Relative to the full model, the vanilla baseline had higher pressure relative-$L_2$ and pressure-drop errors in every stratum (two-sided paired Wilcoxon tests, all $p < 0.001$).

### C.2 Reconstruction-network ablation

Hyperparameters were fixed for all variants and evaluated once on the same frozen 5,000-case test set. Centerline RMSE was calculated from the paired 12-point three-dimensional centerline coordinates, and radius MAE from the corresponding 12 radius values; both are reported in millimetres. Stenosis grading used the prespecified 25%, 50%, and 70% thresholds.

**Table C2-1. Reconstruction-network ablation on the frozen 5,000-case test set. Values are case-wise mean ± population SD.**

| Configuration | Centerline RMSE (mm) | Radius MAE (mm) | Grading error (%) |
|---|---|---|---|
| Full model | 0.288 ± 0.158 | 0.029 ± 0.018 | 7.84 |
| w/o multiscale CBAM | 0.287 ± 0.157 | 0.037 ± 0.024 | 11.42 |
| w/o trainable backbone | 0.356 ± 0.183 | 0.031 ± 0.018 | 8.54 |
| Single-view input, view A | 1.017 ± 0.634 | 0.037 ± 0.024 | 10.36 |
| Single-view input, view B | 1.011 ± 0.632 | 0.039 ± 0.024 | 10.68 |

Removing MS-CBAM did not materially change centerline RMSE: the paired difference, defined as ablation minus full model, was $-0.000391$ mm (10,000-resample bootstrap 95% CI, $-0.002404$ to $0.001665$ mm). In contrast, radius MAE increased by $0.008283$ mm (95% CI, $0.007707$ to $0.008868$ mm), a 28.4% relative increase. Grading errors increased from 392/5,000 (7.84%) to 571/5,000 (11.42%), a difference of 3.58 percentage points (two-sided exact McNemar $p = 4.51 \times 10^{-16}$; Holm-adjusted $p = 3.61 \times 10^{-15}$). Thus, the measurable contribution of the coupled MS-CBAM unit was concentrated in lumen-radius estimation and stenosis grading. Freezing the backbone increased centerline error, whereas either single-view configuration produced the largest centerline deterioration, confirming the importance of trainable dual-view feature extraction.

The full reconstruction model was also compared with models used in Iyer et al. [1] on a common benchmark. Both models used the same 40,000/5,000/5,000 split, two $128 \times 128$ relative-distance-transform views, paired 12-point centerline and radius targets, physical-unit conversion, and case-level evaluation.

**Table C2-2. Common-benchmark results on the frozen 5,000-case test set.**

| Endpoint | Full reconstruction model | Iyer et al. baseline [1] |
|---|---|---|
| Centerline RMSE (mm) | **0.288 ± 0.158** | 0.359 ± 0.146 |
| Radius MAE (mm) | **0.029 ± 0.018** | 0.077 ± 0.035 |
| Stenosis MAE (percentage points) | **1.830** | 7.947 |
| Overall grading error (%) | **7.84** | 36.04 |

Continuous paired differences were estimated using 10,000 case-level bootstrap resamples. The full model reduced centerline RMSE, radius MAE, and stenosis MAE by 19.9%, 62.1%, and 77.0%, respectively.

For context, the original study [1] reported a centerline RMSE of $2.57 \pm 0.78$ mm, an overall radius RMSE of $0.16 \pm 0.07$ mm, and a minimum-stenosis-diameter MAE of $0.27 \pm 0.18$ mm on a different validation set of 500 coronary trees. These published values are not treated as a head-to-head comparison because the image dimensions, view sampling, anatomical outputs, and metric definitions differed from the present benchmark.

### C.3 Measured convergence

Convergence was evaluated on a fixed label-free diagnostic set using the common criterion

$$L_{\mathrm{PDE}} = L_{\mathrm{momentum},x} + L_{\mathrm{momentum},y} + L_{\mathrm{momentum},z} + L_{\mathrm{continuity}} < 3 \times 10^{-6}.$$

The full model reached this criterion in 98/100 cases, compared with 52/100 for the vanilla PINN (Table C3).

**Table C3. Convergence on the fixed 100-case synthetic cohort.**

| Model | Cases reaching $L_{\mathrm{PDE}} < 3 \times 10^{-6}$ | Rate (%) |
|---|---|---|
| JacobiNet–PINN (full model) | 98/100 | 98 |
| Vanilla PINN | 52/100 | 52 |

**Figure C1. Measured convergence comparison for a representative held-out case.**

The cohort-level and case-level results jointly show that the proposed formulation improved convergence robustness while reducing flow-field and pressure-drop errors.

# Appendix-D. Clinical Cohort and Patient-Level Results

## D.1 Patient-specific acquisition geometry and scale harmonization

To provide transparent case-level documentation of the imaging geometry, clinical reference measurements, reconstructed lesion characteristics, and hemodynamic predictions, Table D1 reports the complete acquisition and analysis information for all 32 included patients. For each of the two angiographic views, we provide the recorded positioner angles, including the left/right anterior oblique and cranial/caudal angulations, together with the source-to-image distance (SID), source-to-object distance (SOD), and Imager Pixel Spacing (IPS). The table additionally reports cuff-measured diastolic and systolic blood pressures, invasive FFR, reconstructed diameter stenosis and minimum lumen diameter, predicted pressure drop, and JacobiNet–PINN-derived FFR. All acquisition-geometry and pixel-spacing parameters were extracted directly from the original DICOM metadata and used as recorded. DICOM-to-NIfTI conversion preserved the native pixel data without spatial interpolation or resampling. These case-wise data enable direct traceability from the original image-acquisition geometry and clinical measurements to the reconstructed anatomy and predicted hemodynamic outcomes.

**Table D1. Patient-specific acquisition geometry, clinical measurements, reconstructed lesion geometry, and JacobiNet–PINN results.**

| Study ID | View A angle | View B angle | SID A/B (mm) | SOD A/B (mm) | IPS A/B (mm/pixel) | DBP (mmHg) | SBP (mmHg) | Invasive FFR | Reconstructed DS (%) | Reconstructed MLD (mm) | Predicted pressure drop (Pa) | JacobiNet–PINN FFR |
|---|---|---|---|---|---|---|---|---|---|---|---|---|
| 00001 | LAO 35.80° / CAU 0.14° | RAO 0.95° / CRA 27.40° | 991/ 1098 | 765/ 765 | 0.258391/ 0.258391 | 75 | 121 | 0.878 | 46.71 | 1.518 | 1432.78 | 0.873 |
| 00002 | LAO 44.10° / CAU 1.19° | LAO 0.16° / CRA 28.79° | 1020.1/ 1120.3 | 765/ 765 | 0.305594/ 0.305594 | 94 | 148 | 0.827 | 58.72 | 1.647 | 1908.88 | 0.865 |
| 00003 | LAO 46.40° / CRA 0.17° | RAO 1.65° / CRA 25.40° | 995/ 1122 | 765/ 765 | 0.258391/ 0.258391 | 83 | 115 | 0.810 | 55.06 | 1.537 | 2246.12 | 0.808 |
| 00004 | LAO 43.05° / CAU 2.35° | LAO 4.03° / CRA 26.34° | 990/ 1159 | 765/ 765 | 0.305594/ 0.305594 | 65 | 109 | 0.808 | 58.31 | 1.546 | 2804.18 | 0.714 |
| 00005 | LAO 43.71° / CRA 0.14° | LAO 4.00° / CRA 29.24° | 996/ 1089 | 765/ 765 | 0.305594/ 0.305594 | 65 | 116 | 0.783 | 52.05 | 1.521 | 2074.29 | 0.795 |
| 00006 | LAO 42.52° / CRA 0.86° | LAO 2.21° / CRA 28.86° | 1020/ 1120 | 810/ 810 | 0.305293/ 0.305293 | 82 | 122 | 0.766 | 59.18 | 1.174 | 2648.58 | 0.778 |
| 00007 | LAO 43.80° / CAU 1.83° | LAO 3.25° / CRA 25.11° | 1085/ 1160 | 765/ 765 | 0.258391/ 0.258391 | 96 | 143 | 0.760 | 59.25 | 0.969 | 3193.83 | 0.773 |
| 00008 | LAO 46.40° / CRA 0.43° | RAO 4.45° / CRA 26.31° | 1151/ 1151 | 765/ 765 | 0.258391/ 0.258391 | 76 | 126 | 0.723 | 70.89 | 0.835 | 3650.44 | 0.684 |
| 00009 | LAO 43.50° / CAU 0.72° | RAO 0.44° / CRA 31.20° | 1101/ 1101 | 765/ 765 | 0.258391/ 0.258391 | 74 | 147 | 0.990 | 22.11 | 2.442 | 559.89 | 0.955 |
| 00010 | LAO 44.16° / CAU 1.52° | LAO 3.34° / CRA 30.40° | 1083/ 1096.9 | 809.9/ 810 | 0.305594/ 0.305594 | 80 | 123 | 0.920 | 30.67 | 2.807 | 575.12 | 0.951 |
| 00011 | LAO 43.50° / CAU 0.79° | RAO 3.55° / CRA 29.14° | 1062/ 1083 | 765/ 765 | 0.305594/ 0.305594 | 78 | 131 | 0.940 | 41.01 | 1.707 | 852.33 | 0.929 |
| 00012 | LAO 44.97° / CAU 1.78° | RAO 2.64° / CRA 30.22° | 1078/ 1132 | 765/ 765 | 0.258391/ 0.258391 | 81 | 140 | 0.950 | 40.60 | 1.837 | 924.10 | 0.927 |
| 00013 | LAO 45.08° / CRA 2.96° | RAO 0.14° / CRA 27.94° | 1100/ 1100 | 765/ 765 | 0.305594/ 0.305594 | 67 | 122 | 0.990 | 34.25 | 2.478 | 550.55 | 0.948 |
| 00014 | LAO 46.22° / CAU 4.07° | RAO 2.94° / CRA 26.91° | 1046/ 1046 | 810/ 810 | 0.305293/ 0.305293 | 76 | 106 | 0.970 | 23.11 | 2.659 | 329.02 | 0.969 |
| 00015 | LAO 44.96° / CAU 2.11° | LAO 0.70° / CRA 27.40° | 1178/ 1178 | 765/ 765 | 0.258391/ 0.258391 | 95 | 149 | 0.970 | 45.91 | 1.806 | 861.36 | 0.940 |
| 00016 | LAO 44.61° / CRA 0.02° | RAO 2.29° / CRA 29.05° | 979/ 1028 | 765/ 765 | 0.258391/ 0.258391 | 70 | 126 | 0.930 | 42.68 | 1.671 | 963.49 | 0.913 |
| 00017 | LAO 43.30° / CRA 0.03° | LAO 1.31° / CRA 27.19° | 1037/ 1037 | 765/ 765 | 0.258391/ 0.258391 | 91 | 141 | 0.920 | 44.44 | 1.561 | 891.33 | 0.934 |
| 00018 | LAO 42.74° / CAU 1.31° | LAO 1.75° / CRA 30.76° | 1006/ 1141 | 765/ 765 | 0.258391/ 0.258391 | 76 | 147 | 0.950 | 34.65 | 2.275 | 619.84 | 0.950 |

| Study ID | View A angle | View B angle | SID A/B (mm) | SOD A/B (mm) | IPS A/B (mm/pixel) | DBP (mmHg) | SBP (mmHg) | Invasive FFR | Reconstructed DS (%) | Reconstructed MLD (mm) | Predicted pressure drop (Pa) | JacobiNet–PINN FFR |
|---|---|---|---|---|---|---|---|---|---|---|---|---|
| 00019 | LAO 41.66° / CRA 1.93° | LAO 7.79° / CRA 30.69° | 984/ 1062 | 765/ 765 | 0.258391/ 0.258391 | 61 | 149 | 0.890 | 41.05 | 2.421 | 718.47 | 0.936 |
| 00020 | LAO 43.60° / CAU 10.40° | LAO 2.80° / CRA 28.00° | 1121/ 1183 | 840.137/ 834.961 | 0.154000/ 0.154000 | 61 | 102 | 0.860 | 50.45 | 1.679 | 1571.11 | 0.828 |
| 00021 | LAO 37.41° / CAU 0.96° | LAO 3.48° / CRA 17.32° | 958/ 1042 | 765/ 765 | 0.258391/ 0.258391 | 72 | 118 | 0.660 | 59.08 | 1.207 | 3481.39 | 0.679 |
| 00022 | LAO 43.09° / CAU 1.03° | RAO 4.49° / CRA 25.88° | 1103/ 1173 | 765/ 765 | 0.258391/ 0.258391 | 56 | 106 | 0.720 | 54.65 | 1.402 | 2815.54 | 0.683 |
| 00023 | LAO 45.34° / CRA 0.24° | RAO 5.82° / CRA 21.41° | 947/ 1007 | 765/ 765 | 0.258391/ 0.258391 | 58 | 105 | 0.830 | 59.91 | 1.153 | 2204.76 | 0.756 |
| 00024 | LAO 40.52° / CAU 2.91° | RAO 1.09° / CRA 33.09° | 983/ 1063 | 765/ 765 | 0.258391/ 0.258391 | 62 | 140 | 0.870 | 44.14 | 1.757 | 1310.73 | 0.880 |
| 00025 | LAO 43.69° / CRA 1.02° | LAO 9.67° / CRA 35.84° | 1044/ 1215 | 765/ 765 | 0.305594/ 0.305594 | 82 | 128 | 0.900 | 40.21 | 2.094 | 953.12 | 0.922 |
| 00026 | LAO 45.70° / CAU 1.60° | RAO 1.50° / CRA 30.10° | 1070/ 1070 | 724.088/ 738.904 | 0.154000/ 0.154000 | 70 | 112 | 0.800 | 52.63 | 1.238 | 2340.66 | 0.775 |
| 00027 | LAO 45.48° / CRA 0.28° | LAO 7.76° / CRA 30.22° | 1027/ 962 | 765/ 765 | 0.305594/ 0.305594 | 80 | 121 | 0.920 | 43.52 | 1.702 | 1088.28 | 0.907 |
| 00028 | LAO 45.00° / CRA 1.80° | LAO 3.40° / CRA 27.60° | 1045/ 1135 | 765/ 765 | 0.258391/ 0.258391 | 82 | 122 | 0.740 | 55.96 | 1.443 | 3410.28 | 0.714 |
| 00029 | LAO 38.30° / CAU 0.60° | RAO 5.80° / CRA 25.00° | 1103/ 1069 | 791.077/ 787.210 | 0.140000/ 0.140000 | 79 | 143 | 0.940 | 39.96 | 1.723 | 904.69 | 0.928 |
| 00030 | LAO 43.50° / CRA 0.25° | LAO 0.14° / CRA 29.81° | 1105/ 1144 | 765/ 765 | 0.258391/ 0.258391 | 71 | 151 | 0.860 | 56.22 | 1.252 | 1884.86 | 0.846 |
| 00031 | LAO 45.18° / CAU 2.07° | LAO 0.66° / CRA 29.79° | 1006/ 1053 | 765/ 765 | 0.258391/ 0.258391 | 53 | 102 | 0.730 | 54.36 | 1.526 | 2784.30 | 0.670 |
| 00032 | LAO 46.90° / CRA 0.80° | LAO 2.10° / CRA 30.50° | 1200/ 1140 | 842.111/ 825.875 | 0.140000/ 0.140000 | 74 | 113 | 0.960 | 34.91 | 1.727 | 900.81 | 0.917 |

*A/B denote the two selected projections. Directional labels follow the DICOM detector-positioner convention. Because the synthetic generator specifies the opposing X-ray-source direction, LAO/CRA detector positions correspond to RAO/CAU source-side configurations.*

### D.2 Diagnostic performance and agreement with measured FFR

The prespecified positive-class definition was invasive FFR $\leq 0.80$. 9 patients were positive and 23 were negative. JacobiNet–PINN yielded 9 true positives, no false negative, 21 true negatives, and 2 false positives (Table D2-1).

**Table D2-1. Confusion matrix for JacobiNet–PINN FFR.**

| | Measured positive | Measured negative | Total |
|---|---|---|---|
| JacobiNet–PINN positive | 9 | 2 | 11 |
| JacobiNet–PINN negative | 0 | 21 | 21 |
| Total | 9 | 23 | 32 |

**Table D2-2. Diagnostic performance at the measured-FFR threshold of 0.80.**

| Metric | Estimate | 95% CI |
|---|---|---|
| Sensitivity | 100.0% | 66.4%–100.0% |
| Specificity | 91.3% | 72.0%–98.9% |
| Positive predictive value | 81.8% | 48.2%–97.7% |
| Negative predictive value | 100.0% | 83.9%–100.0% |
| Accuracy | 93.8% (30/32) | 79.2%–99.2% |
| Balanced accuracy | 95.7% | — |
| AUC | 0.961 | 0.884–1.000 |

Confidence intervals for sensitivity, specificity, predictive values, and accuracy are two-sided Clopper–Pearson intervals. The AUC interval was obtained by 10,000 stratified patient-level bootstrap resamples. An all-negative classifier would have achieved 71.9% accuracy in this cohort; the positive-class and predictive-value intervals remain wide because only nine patients were positive.

Continuous agreement was evaluated without using Pearson correlation as an agreement statistic. The patient was the sampling unit for all confidence intervals.

**Table D2-3. Continuous agreement between JacobiNet–PINN and invasive FFR.**

| Metric | Result |
|---|---|
| Invasive FFR | 0.862 ± 0.091 |
| JacobiNet–PINN FFR | 0.848 ± 0.098 |
| Mean difference, JacobiNet–PINN minus invasive FFR | −0.013 |
| Difference SD | 0.032 |
| Mean-difference 95% CI | −0.025 to −0.002 |
| Mean absolute error | 0.027 |
| Root-mean-square error | 0.034 |
| Lin’s concordance correlation coefficient | 0.934 |
| Lin’s CCC 95% CI | 0.885–0.964 |
| Bland–Altman 95% limits of agreement | −0.076 to 0.049 |
| Lower-limit 95% CI | −0.096 to −0.056 |
| Upper-limit 95% CI | 0.029–0.069 |

| Metric | Result |
|---|---|
| Proportional-bias slope | 0.074 |
| Slope 95% CI | −0.050 to 0.198 |
| Proportional-bias test | $p = 0.235$ |

The Bland–Altman difference was defined as JacobiNet–PINN FFR minus invasive FFR. The proportional-bias slope was obtained by regressing the paired difference on the paired mean. No statistically detectable proportional bias was observed, although the limits of agreement require cautious interpretation for individual predictions near the 0.80 threshold.

## D.3 Patient-level full-field errors under repeated flow conditions

Each patient was evaluated at four inlet-flow conditions. The 128 patient–condition evaluations are repeated observations nested within 32 patients and were not treated as 128 independent samples. Table D3-1 summarizes each condition across patients, and Table D3-2 reports all patient-level values. Each entry in Table D3-2 is velocity relative-$L_2$/pressure relative-$L_2$.

**Table D3-1. Condition-specific PINN–CFD full-field errors.**

| Mean inlet velocity (m/s) | Patients, $n$ | Velocity relative-$L_2$ | Pressure relative-$L_2$ |
|---|---|---|---|
| 0.144 | 32 | 0.042 ± 0.015 | 0.008 ± 0.003 |
| 0.169 | 32 | 0.044 ± 0.018 | 0.012 ± 0.009 |
| 0.209 | 32 | 0.050 ± 0.024 | 0.018 ± 0.021 |
| 0.383 | 32 | 0.080 ± 0.037 | 0.053 ± 0.039 |

After first averaging the four conditions within each patient, the patient-level means were $0.054 \pm 0.023$ for velocity and $0.023 \pm 0.016$ for pressure.

**Table D3-2. Patient-level velocity/pressure relative-$L_2$ errors.**

| Study ID | 0.144 m/s | 0.169 m/s | 0.209 m/s | 0.383 m/s | Four-condition mean |
|---|---|---|---|---|---|
| 00001 | 0.036/0.005 | 0.035/0.012 | 0.038/0.028 | 0.069/0.109 | 0.045/0.039 |
| 00002 | 0.038/0.008 | 0.039/0.007 | 0.044/0.017 | 0.079/0.082 | 0.050/0.028 |
| 00003 | 0.043/0.011 | 0.045/0.009 | 0.049/0.007 | 0.074/0.036 | 0.053/0.016 |
| 00004 | 0.052/0.006 | 0.067/0.035 | 0.088/0.039 | 0.165/0.071 | 0.093/0.038 |
| 00005 | 0.067/0.012 | 0.070/0.011 | 0.073/0.010 | 0.077/0.011 | 0.072/0.011 |
| 00006 | 0.046/0.010 | 0.048/0.007 | 0.053/0.007 | 0.083/0.037 | 0.057/0.015 |
| 00007 | 0.049/0.008 | 0.051/0.020 | 0.057/0.038 | 0.101/0.120 | 0.065/0.047 |
| 00008 | 0.090/0.012 | 0.102/0.017 | 0.118/0.024 | 0.165/0.058 | 0.119/0.027 |
| 00009 | 0.027/0.005 | 0.027/0.005 | 0.027/0.005 | 0.027/0.010 | 0.027/0.007 |
| 00010 | 0.027/0.006 | 0.027/0.006 | 0.027/0.009 | 0.042/0.043 | 0.031/0.016 |
| 00011 | 0.032/0.010 | 0.033/0.009 | 0.034/0.008 | 0.046/0.026 | 0.036/0.013 |
| 00012 | 0.032/0.010 | 0.031/0.008 | 0.031/0.007 | 0.051/0.038 | 0.036/0.016 |
| 00013 | 0.031/0.011 | 0.031/0.009 | 0.033/0.008 | 0.054/0.032 | 0.037/0.015 |

| Study ID | 0.144 m/s | 0.169 m/s | 0.209 m/s | 0.383 m/s | Four-condition mean |
|---|---|---|---|---|---|
| 00014 | 0.027/0.006 | 0.026/0.007 | 0.026/0.007 | 0.030/0.018 | 0.027/0.010 |
| 00015 | 0.036/0.014 | 0.037/0.013 | 0.041/0.012 | 0.070/0.014 | 0.046/0.013 |
| 00016 | 0.034/0.011 | 0.033/0.010 | 0.034/0.009 | 0.054/0.029 | 0.039/0.015 |
| 00017 | 0.035/0.010 | 0.035/0.010 | 0.035/0.008 | 0.053/0.026 | 0.039/0.013 |
| 00018 | 0.029/0.008 | 0.028/0.009 | 0.027/0.009 | 0.042/0.012 | 0.032/0.009 |
| 00019 | 0.030/0.008 | 0.030/0.007 | 0.031/0.010 | 0.054/0.048 | 0.036/0.018 |
| 00020 | 0.034/0.009 | 0.034/0.007 | 0.037/0.005 | 0.064/0.026 | 0.042/0.012 |
| 00021 | 0.073/0.007 | 0.081/0.014 | 0.094/0.025 | 0.137/0.071 | 0.096/0.029 |
| 00022 | 0.042/0.007 | 0.048/0.009 | 0.068/0.024 | 0.153/0.089 | 0.078/0.032 |
| 00023 | 0.043/0.007 | 0.045/0.006 | 0.048/0.007 | 0.069/0.024 | 0.051/0.011 |
| 00024 | 0.037/0.011 | 0.038/0.009 | 0.042/0.011 | 0.067/0.045 | 0.046/0.019 |
| 00025 | 0.034/0.011 | 0.032/0.010 | 0.034/0.011 | 0.078/0.049 | 0.044/0.020 |
| 00026 | 0.046/0.006 | 0.052/0.053 | 0.078/0.117 | 0.097/0.134 | 0.068/0.077 |
| 00027 | 0.041/0.006 | 0.042/0.017 | 0.046/0.035 | 0.085/0.118 | 0.053/0.044 |
| 00028 | 0.052/0.006 | 0.055/0.012 | 0.061/0.024 | 0.106/0.079 | 0.068/0.030 |
| 00029 | 0.043/0.011 | 0.044/0.010 | 0.047/0.008 | 0.066/0.015 | 0.050/0.011 |
| 00030 | 0.037/0.004 | 0.038/0.017 | 0.046/0.040 | 0.107/0.158 | 0.057/0.055 |
| 00031 | 0.071/0.009 | 0.085/0.007 | 0.101/0.010 | 0.139/0.040 | 0.099/0.016 |
| 00032 | 0.034/0.006 | 0.033/0.006 | 0.033/0.006 | 0.073/0.040 | 0.043/0.015 |

### D.4 Collocation sampling and case-wise point counts

For each patient-specific geometry, the interior and boundary point sets were generated once and remained fixed throughout training. A seeded sampler traversed randomized permutations of the interior PDE pool in batches of 8,192, without replacement or omission within each pass, and reshuffled only after the pool had been exhausted. Because the inlet, outlet, and wall conditions were analytically encoded through JacobiNet and the trial functions, boundary points were excluded from optimization and used only for evaluation. A separate fixed set of 2,048 interior points was used for label-free validation.

Across the 32 geometries, the interior PDE pools contained 20,849–161,710 points (mean, 57,118), and the combined inlet, outlet, and wall sets contained 10,646–41,483 points (mean, 19,694). The case-wise counts are reported in Table D4.

**Table D4. Fixed point-set sizes for the 32 patient-specific geometries.**

| Study ID | Interior PDE points | Boundary evaluation points |
|---|---|---|
| 00001 | 50,095 | 20,177 |
| 00002 | 124,595 | 35,951 |
| 00003 | 70,427 | 24,055 |
| 00004 | 64,103 | 21,059 |
| 00005 | 51,610 | 18,576 |
| 00006 | 34,581 | 14,844 |

| Study ID | Interior PDE points | Boundary evaluation points |
| --- | --- | --- |
| 00007 | 20,849 | 10,646 |
| 00008 | 35,929 | 15,741 |
| 00009 | 61,832 | 20,569 |
| 00010 | 161,710 | 41,483 |
| 00011 | 27,120 | 11,486 |
| 00012 | 49,715 | 18,187 |
| 00013 | 83,530 | 23,821 |
| 00014 | 82,802 | 24,926 |
| 00015 | 40,178 | 14,392 |
| 00016 | 37,329 | 14,742 |
| 00017 | 36,667 | 15,024 |
| 00018 | 70,179 | 22,098 |
| 00019 | 91,052 | 25,034 |
| 00020 | 60,236 | 20,088 |
| 00021 | 42,997 | 17,411 |
| 00022 | 63,557 | 23,915 |
| 00023 | 30,447 | 12,704 |
| 00024 | 35,864 | 13,584 |
| 00025 | 87,534 | 27,795 |
| 00026 | 28,180 | 12,920 |
| 00027 | 59,591 | 22,089 |
| 00028 | 36,817 | 14,037 |
| 00029 | 38,914 | 15,194 |
| 00030 | 50,828 | 20,039 |
| 00031 | 60,187 | 21,376 |
| 00032 | 38,329 | 16,258 |

## Appendix-E. Clinical Hemodynamic Controls, Transfer Learning, and Computational Efficiency

This appendix reports the clinical hemodynamic controls and sensitivity analyses performed in the 32-patient RCA cohort. Unless otherwise stated, values are patient-level means $\pm$ sample standard deviations, and FFR classification uses invasive FFR $\leq 0.80$ as the positive reference.

### E.1 Reynolds-number range and physics-only transfer learning

For each patient and flow condition, the inlet and stenosis-throat Reynolds numbers were calculated as

$$Re_{\text{in}} = \frac{\rho U_{\text{in}} D_{\text{in}}}{\mu}, Re_{\text{th}} = \frac{\rho U_{\text{th}} D_{\text{th}}}{\mu},$$

where $\rho = 1060$ kg/m$^3$ and $\mu = 0.004$ Pa·s. The four evaluated mean inlet velocities covered inlet Reynolds numbers of 90.6–415.9 and throat Reynolds numbers of 152.5–871.6 (Table E1).

**Table E1-1. Reynolds-number distributions across the 32 clinical geometries.**

| Mean inlet velocity (m/s) | Inlet $Re$, mean $\pm$ SD (range) | Throat $Re$, mean $\pm$ SD (range) |
|---|---|---|
| 0.144 | 121.4 $\pm$ 16.0 (90.6–156.4) | 229.2 $\pm$ 47.8 (152.5–327.7) |
| 0.169 | 142.5 $\pm$ 18.8 (106.4–183.5) | 269.0 $\pm$ 56.1 (179.0–384.6) |
| 0.209 | 176.2 $\pm$ 23.3 (131.6–226.9) | 332.6 $\pm$ 69.3 (221.3–475.6) |
| 0.383 | 322.9 $\pm$ 42.7 (241.1–415.9) | 609.5 $\pm$ 127.1 (405.6–871.6) |

**Table E1-2. Case-wise inlet/throat Reynolds numbers.**

| Study ID | 0.144 m/s | 0.169 m/s | 0.209 m/s | 0.383 m/s |
|---|---|---|---|---|
| 00001 | 108.9 / 194.5 | 127.8 / 228.3 | 158.1 / 282.3 | 289.7 / 517.4 |
| 00002 | 152.2 / 327.7 | 178.6 / 384.6 | 220.9 / 475.6 | 404.7 / 871.6 |
| 00003 | 130.4 / 283.1 | 153.0 / 332.3 | 189.2 / 411.0 | 346.7 / 753.1 |
| 00004 | 141.5 / 327.2 | 166.1 / 384.1 | 205.4 / 475.0 | 376.4 / 870.4 |
| 00005 | 118.9 / 255.6 | 139.6 / 300.0 | 172.6 / 371.0 | 316.3 / 679.9 |
| 00006 | 109.4 / 246.4 | 128.4 / 289.2 | 158.8 / 357.7 | 290.9 / 655.5 |
| 00007 | 90.6 / 211.5 | 106.4 / 248.2 | 131.6 / 307.0 | 241.1 / 562.5 |
| 00008 | 109.1 / 267.8 | 128.0 / 314.3 | 158.3 / 388.7 | 290.1 / 712.3 |
| 00009 | 119.6 / 152.5 | 140.4 / 179.0 | 173.7 / 221.3 | 318.2 / 405.6 |
| 00010 | 153.9 / 212.1 | 180.7 / 249.0 | 223.4 / 307.9 | 409.5 / 564.2 |
| 00011 | 110.5 / 181.2 | 129.6 / 212.7 | 160.3 / 263.0 | 293.8 / 482.0 |
| 00012 | 118.1 / 187.8 | 138.5 / 220.4 | 171.3 / 272.6 | 314.0 / 499.6 |
| 00013 | 143.8 / 226.0 | 168.8 / 265.2 | 208.7 / 328.0 | 382.5 / 601.0 |
| 00014 | 131.8 / 169.9 | 154.7 / 199.4 | 191.3 / 246.6 | 350.5 / 451.8 |
| 00015 | 127.4 / 224.0 | 149.5 / 262.9 | 184.9 / 325.2 | 338.8 / 595.9 |
| 00016 | 111.1 / 184.4 | 130.3 / 216.4 | 161.2 / 267.6 | 295.4 / 490.3 |

| Study ID | 0.144 m/s | 0.169 m/s | 0.209 m/s | 0.383 m/s |
|---|---|---|---|---|
| 00017 | 106.9 / 182.0 | 125.5 / 213.7 | 155.2 / 264.2 | 284.4 / 484.2 |
| 00018 | 132.8 / 198.7 | 155.9 / 233.1 | 192.8 / 288.3 | 353.3 / 528.4 |
| 00019 | 156.4 / 257.4 | 183.5 / 302.1 | 226.9 / 373.6 | 415.9 / 684.6 |
| 00020 | 129.2 / 246.1 | 151.6 / 288.8 | 187.5 / 357.1 | 343.6 / 654.5 |
| 00021 | 112.5 / 288.0 | 132.1 / 338.1 | 163.3 / 418.1 | 299.3 / 766.1 |
| 00022 | 118.0 / 260.3 | 138.5 / 305.5 | 171.3 / 377.8 | 313.8 / 692.4 |
| 00023 | 112.4 / 244.8 | 131.9 / 287.3 | 163.1 / 355.3 | 298.9 / 651.1 |
| 00024 | 120.1 / 226.2 | 140.9 / 265.5 | 174.2 / 328.3 | 319.3 / 601.7 |
| 00025 | 133.7 / 227.4 | 156.9 / 266.9 | 194.0 / 330.1 | 355.5 / 604.9 |
| 00026 | 99.7 / 213.1 | 117.0 / 250.1 | 144.8 / 309.2 | 265.3 / 566.7 |
| 00027 | 114.9 / 190.9 | 134.8 / 224.1 | 166.7 / 277.1 | 305.6 / 507.9 |
| 00028 | 124.3 / 296.8 | 145.9 / 348.3 | 180.4 / 430.7 | 330.6 / 789.3 |
| 00029 | 109.1 / 168.7 | 128.0 / 198.0 | 158.3 / 244.9 | 290.1 / 448.7 |
| 00030 | 109.4 / 226.8 | 128.4 / 266.2 | 158.7 / 329.2 | 290.9 / 603.3 |
| 00031 | 126.7 / 297.6 | 148.7 / 349.3 | 183.9 / 431.9 | 337.0 / 791.5 |
| 00032 | 101.2 / 156.9 | 118.8 / 184.1 | 146.9 / 227.7 | 269.3 / 417.3 |

Values are reported as inlet $Re$ / throat $Re$.

The 0.144-m/s solution was used as the source condition. For each target velocity, $\alpha = U_{\text{target}}/0.144$. Deterministic controls were formed as $\boldsymbol{u}_{\text{lin}} = \alpha \boldsymbol{u}_{0.144}$, $p_\alpha = \alpha p_{0.144}$, and $p_{\alpha^2} = \alpha^2 p_{0.144}$. The transferred model was initialized from the preceding-condition network and fine-tuned for 1,000 physics-only steps without target-condition CFD labels.

**Table E1-3. Physics-only transfer learning versus deterministic field scaling.**

| Target velocity | Velocity $L_2$: linear $\alpha_u$ | Velocity $L_2$: transfer | Improvement vs $\alpha_u$ | Pressure $L_2$: $\alpha_p$ | Pressure $L_2$: $\alpha_p^2$ | Pressure $L_2$: transfer | Improvement vs $\alpha_p$ | Improvement vs $\alpha_p^2$ |
|---|---|---|---|---|---|---|---|---|
| 0.169 m/s | $0.048 \pm 0.017$ | $0.044 \pm 0.018$ | 6.53% | $0.099 \pm 0.017$ | $0.059 \pm 0.020$ | $0.012 \pm 0.009$ | 87.81% | 79.52% |
| 0.209 m/s | $0.067 \pm 0.017$ | $0.050 \pm 0.024$ | 25.62% | $0.222 \pm 0.033$ | $0.131 \pm 0.050$ | $0.018 \pm 0.021$ | 91.78% | 86.01% |
| 0.383 m/s | $0.135 \pm 0.025$ | $0.080 \pm 0.037$ | 40.43% | $0.506 \pm 0.053$ | $0.319 \pm 0.148$ | $0.053 \pm 0.039$ | 89.45% | 83.26% |

Transfer learning reduced velocity error in 26/32, 28/32, and 28/32 patients and pressure error relative to linear scaling in 32/32 patients at the three target velocities. Mean pressure error relative to the $\alpha^2$ control decreased by 79.52%, 86.01%, and 83.26%, respectively. Across the 96 transfers, mean wall-clock time was $13.79 \pm 2.07$ s per target condition.

## E.2 Sensitivity to the standardized inlet velocity

The nominal hyperemic mean inlet velocity of 0.383 m/s was prespecified from an RCA whole-cycle velocity of 38.3 cm/s and was not fitted to the present cohort [18]. Six additional conditions spanning $-50\%$ to $+50\%$ were evaluated while holding geometry and all other settings fixed.

**Table E2. Patient-level sensitivity to inlet velocity.**

| Relative velocity | Inlet velocity (m/s) | JacobiNet–PINN FFR | Paired $\Delta$FFR vs 0.383 m/s | FFR accuracy (%) |
|---|---|---|---|---|
| 50% | 0.1915 | $0.934 \pm 0.036$ | $+0.086 \pm 0.063$ | 71.9 |
| 70% | 0.2680 | $0.903 \pm 0.060$ | $+0.054 \pm 0.038$ | 87.5 |
| 90% | 0.3447 | $0.867 \pm 0.083$ | $+0.019 \pm 0.015$ | 90.6 |
| 100% | 0.3830 | $0.848 \pm 0.098$ | 0 | 93.8 |
| 110% | 0.4213 | $0.824 \pm 0.115$ | $-0.024 \pm 0.019$ | 93.8 |
| 130% | 0.4979 | $0.776 \pm 0.151$ | $-0.072 \pm 0.055$ | 87.5 |
| 150% | 0.5745 | $0.718 \pm 0.194$ | $-0.130 \pm 0.099$ | 75.0 |

Accuracy remained 87.5%–93.8% within the $\pm30\%$ perturbation range. The $\pm50\%$ conditions were included as conservative outer-tail stress tests and produced larger, directionally coherent FFR changes.

## E.3 Geometric and analytical baselines

Diameter stenosis (DS) and minimal lumen diameter (MLD) were calculated from the unprocessed reconstruction-network output. The Young–Tsai analytical law was evaluated on the same final processed geometries and at the same 0.383-m/s inlet velocity used by CFD and JacobiNet–PINN, with $K_t = 1.52$ and $L_a = 0.83L_s + 1.64D_s$ [19–22].

**Table E3. Diagnostic performance of geometric, analytical, and JacobiNet–PINN methods.**

| Method | Prespecified threshold | Accuracy | FFR MAE | FFR RMSE | AUC |
|---|---|---|---|---|---|
| Diameter stenosis | DS > 43.5% [23] | 65.6% (21/32) | — | — | 0.870 |
| | DS ≥ 48% [24] | 81.3% (26/32) | — | — | 0.870 |
| | DS > 50% [25] | 81.3% (26/32) | — | — | 0.870 |
| | DS ≥ 52% [25] | 84.4% (27/32) | — | — | 0.870 |
| Minimal lumen diameter | MLD <1.31 mm [26] | 81.3% (26/32) | — | — | 0.937 |
| | MLD <1.35 mm [23] | 81.3% (26/32) | — | — | 0.937 |
| | MLD ≤ 1.50 mm [24] | 87.5% (28/32) | — | — | 0.937 |
| Young–Tsai analytical law | FFR ≤ 0.80 | 90.6% (29/32) | 0.045 | 0.063 | **0.966** |
| **JacobiNet–PINN** | FFR ≤ 0.80 | **93.8% (30/32)** | **0.027** | **0.034** | 0.961 |

JacobiNet–PINN achieved the highest diagnostic accuracy and the lowest continuous FFR error. Against CFD, the Young–Tsai law yielded a pressure-drop MAE of 199.06 Pa, RMSE of 322.96 Pa, and mean absolute percentage error of 12.263%; the corresponding JacobiNet–PINN values were 103.25 Pa, 170.26 Pa, and 5.435%. The Young–Tsai law remained competitive in rank discrimination but did not provide spatially resolved velocity, pressure, or WSS fields.

### E.4 Patient-specific aortic-pressure normalization

Two pressure-normalization schemes were compared: a fixed reference of 100 mmHg and patient-specific hyperemic mean aortic pressure estimated from the recorded cuff systolic and diastolic pressures:

$$P_{a,i}^{\mathrm{MAP}} = \frac{\mathrm{SBP}_i + 2\mathrm{DBP}_i}{3} - 6\ \mathrm{mmHg}, \mathrm{FFR}_{\mathrm{MAP},i} = 1 - \frac{\Delta P_i}{133.322 P_{a,i}^{\mathrm{MAP}}}.$$

The 6-mmHg correction represents the mean aortic-pressure reduction reported during adenosine-induced maximal hyperemia [27].

**Table E4. Fixed-pressure and patient-specific MAP normalization.**

| Metric | Fixed 100 mmHg | Patient-specific MAP |
|---|---|---|
| Normalization pressure (mmHg) | 100 | 85.81 ± 11.11 (63.33–107.00) |
| JacobiNet–PINN FFR | 0.875 ± 0.075 | 0.848 ± 0.098 |
| FFR MAE | 0.025 | 0.027 |
| FFR RMSE | 0.033 | 0.034 |
| AUC (95% CI) | 0.971 (0.908–1.000) | 0.961 (0.884–1.000) |
| Accuracy | 87.5% (28/32) | 93.8% (30/32) |
| Sensitivity | 66.7% (6/9) | 100.0% (9/9) |
| Specificity | 95.7% (22/23) | 91.3% (21/23) |

Patient-specific normalization improved accuracy, particularly sensitivity, while the small increases in FFR MAE and RMSE remained limited.

### E.5 Distal pressure-wire position

The retrospective angiographic records contained the angiographic images but not calibrated three-dimensional pressure-sensor coordinates. The additional unmodeled segment between the reconstructed outlet and the pressure-readout plane was therefore represented by a prespecified length of 15 mm [28]. Its viscous loss was calculated as

$$\Delta P_{\mathrm{distal}} = \frac{64}{Re_{\mathrm{outlet}}} \frac{L}{D_{\mathrm{outlet}}} \frac{\rho \bar{v}_{\mathrm{outlet}}^2}{2}, Re_{\mathrm{outlet}} = \frac{\rho \bar{v}_{\mathrm{outlet}} D_{\mathrm{outlet}}}{\mu},$$

with $L = 15$ mm and the case-specific outlet diameter and mean velocity.

**Table E5. Effect of the additional outlet-to-sensor pressure loss.**

| Metric | Result |
|---|---|
| Prespecified distal length | 15 mm |
| Additional distal pressure loss | 115.60 ± 32.82 Pa |
| Absolute FFR change | 0.0102 ± 0.0030 |
| Maximum absolute FFR change | 0.0161 |
| Threshold crossings | 1/32 |

The single threshold crossing occurred in a near-threshold patient in both the CFD and JacobiNet–PINN analyses; the cohort-level conclusion was unchanged.

### E.6 Reduced main-vessel through-flow

Because distal RCA tapering can partly preserve mean velocity despite side-branch flow loss [29–31], this experiment evaluates reduced flow through the modeled lesion rather than explicit branch-resolved hemodynamics. Mean inlet velocity was reduced by 10% and 30% while inlet area, geometry, and all other settings were fixed.

**Table E6. Sensitivity to reduced lesion through-flow.**

| Prescribed reduction | Mean inlet velocity (m/s) | JacobiNet–PINN FFR | FFR accuracy (%) |
|---|---|---|---|
| Baseline | 0.3830 | $0.848 \pm 0.098$ | 93.8 |
| 10% | 0.3447 | $0.867 \pm 0.083$ | 90.6 |
| 30% | 0.2680 | $0.903 \pm 0.060$ | 87.5 |

Lower lesion through-flow progressively increased FFR, with mean changes of 0.019 and 0.054 at the 10% and 30% reductions. Accuracy remained at least 87.5%; changes were concentrated in threshold-adjacent positive cases.

### E.7 Wall-clock efficiency

All timings were recorded on the same workstation. Manual segmentation, dual-view reconstruction, and virtual-geometry generation are shared preprocessing stages and are counted once. The 32-patient four-condition analysis comprised 128 CFD runs; all met the prespecified residual and mass-balance criteria. Virtual-revascularization timing was evaluated in the nine patients with invasive FFR $\leq 0.80$.

**Table E7. Stage-wise wall-clock time.**

| Workflow stage | JacobiNet–PINN workflow, s (mean ± SD) | Independent CFD workflow, s (mean ± SD) |
|---|---|---|
| Manual image segmentation | $67.62 \pm 14.68$ | |
| Dual-view images to three-dimensional lumen geometry | $1.68 \pm 0.27$ | |
| Geometry-specific preparation | JacobiNet training: $43.41 \pm 20.74$ | Mesh generation: $53.20 \pm 6.73$ |
| Initial condition: 0.144 m/s | $199.84 \pm 59.92$ | $224.89 \pm 81.54$ |
| Additional condition: 0.169 m/s | $13.82 \pm 1.92$ | $225.69 \pm 82.87$ |
| Additional condition: 0.209 m/s | $13.89 \pm 1.96$ | $224.45 \pm 83.90$ |
| Additional condition: 0.383 m/s | $13.73 \pm 2.35$ | $206.19 \pm 78.22$ |
| Generation of three virtual-revascularization geometries | $4.74 \pm 0.49$ | |
| Hemodynamic evaluation of three virtual-revascularization candidates | JacobiNet and PINN training: $623.64 \pm 66.46$ | CFD meshing and solves: $1{,}113.80 \pm 237.29$ |
| Complete end-to-end total including virtual revascularization | $1{,}030.88 \pm 40.95$ | $1{,}994.86 \pm 512.47$ |

For a single condition, JacobiNet–PINN and CFD required comparable wall-clock time. Reuse through physics-only transfer reduced each additional-condition calculation to 13.73–13.89 s, compared with 206.19–225.69 s for an independent CFD solve.

### E.8 Context relative to established functional-imaging methods

Cross-study values in Table E8 are descriptive because populations, reference standards, endpoints, and validation levels differ. Outcome studies of stress CMR were not treated as lesion-level diagnostic-accuracy studies.

**Table E8. Descriptive comparison with representative functional-assessment methods.**

| Approach and representative study | Imaging input | Patients scale | Reported diagnostic evidence | Reported analysis time (s) | Primary output / clinical applicability |
|---|---|---|---|---|---|
| $FFR_{QCA}$ [32] | Coronary angiography | 68 | Accuracy 88.3%; AUC 0.93 | Approximately 300 | FFR estimate; longitudinal pullback |
| QFR (FAVOR II China) [33] | Coronary angiography | 308 | Accuracy 92.7% (95% CI, 0.89–0.95) | $261.6 \pm 153.0$ | FFR estimate; longitudinal pullback |
| vFFR (FAST/FAST II) [34,35] | Coronary angiography | 100/334 | Accuracy 90%; AUC 0.93 (95% CI, 0.90–0.96) | Not reported | FFR estimate; longitudinal pullback |
| caFFR (FLASH FFR) [36] | Coronary angiography | 328 | Accuracy 95.7%; AUC 0.979 | $272.4 \pm 88.8$ | FFR estimate |
| $\mu$QFR [37] | Coronary angiography | 306 | Accuracy 93.0% (95% CI, 0.90–0.96) | $67 \pm 22$ | FFR estimate; longitudinal pullback |
| Photon-counting CT-FFR [38,39] | Photon-counting CCTA | 32 | SR-PCCTA AUC 0.80; UHR-PCCTA AUC 0.93 | Not reported | FFR estimate |
| Stress perfusion CMR [40] | Stress perfusion CMR | 7,113 | Sensitivity 0.90; specificity 0.85 | Not reported | Myocardial perfusion; no lesion-level FFR |
| **JacobiNet–PINN (present study)** | **Coronary angiography** | **32** | **Accuracy 93.8%; AUC 0.961** | $353.99 \pm 61.32$ | **FFR estimate, longitudinal pullback, three-dimensional velocity and pressure fields, and WSS distribution** |

The present workflow time includes manual segmentation, dual-view reconstruction, geometry-specific preparation, initial training at 0.144 m/s, and transfer to 0.169, 0.209, and 0.383 m/s. Its principal distinction is the recovery of spatially resolved three-dimensional fields and their efficient reuse across conditions; Table E8 is not a head-to-head diagnostic comparison.

## Appendix-F. Reconstruction Validation and Robustness

This appendix evaluates reconstruction robustness to projection-angle variation, segmentation uncertainty, observer variability, eccentric stenosis, and non-circular cross-sections. The projection-angle experiment used the fixed 100-case synthetic cohort; clinical robustness analyses used the 32-patient RCA cohort.

### F.1 Projection-angle sensitivity

The reconstruction network was trained using two complementary RCA views with bounded angular variation. In the X-ray-source-side convention, the first view was sampled at RAO $40^\circ \pm 10^\circ$ with CRA/CAU $0^\circ \pm 5^\circ$; the second was sampled within $5^\circ$ of RAO/LAO $0^\circ$ with CAU $25^\circ \pm 10^\circ$. The angular components were sampled independently from bounded uniform distributions. SID, SOD, and detector pixel spacing were fixed at 1,100 mm, 765 mm, and 0.258390625 mm/pixel, respectively. These complementary views are consistent with standard RCA angiographic practice [41–43]. Patient-specific acquisition geometry is reported in Appendix D.

Each synthetic case's original view pair was retained as the 0° reference. The two viewing directions were displaced symmetrically along their shared great circle using total signed perturbations of $\pm 5^\circ$, $\pm 10^\circ$, $\pm 20^\circ$, and $\pm 40^\circ$. Both projections were regenerated at each condition and processed by the same frozen reconstruction model. SID, SOD, pixel spacing, preprocessing, and model weights were unchanged. The reconstructed geometries were evaluated against the labeled geometry and then analyzed using the same meshing protocol and 0.383-m/s CFD. Because no independent FFR reference existed for the perturbed reconstructions, downstream sensitivity was measured as the paired absolute FFR deviation from the same case's 0° result.

**Table F1. Projection-angle sensitivity in the fixed 100-case synthetic cohort.**

| Total signed paired-view perturbation | Inter-view angle (°) | Centerline $RMSE_{xyz}$ (mm) | Radius $MAE_r$ (mm) | Paired absolute FFR deviation from 0° |
|---|---|---|---|---|
| $-40^\circ$ | $7.6726 \pm 4.5168$ | $1.4506 \pm 0.8814$ | $0.0358 \pm 0.0217$ | $0.1633 \pm 0.3842$ |
| $-20^\circ$ | $27.2295 \pm 5.1967$ | $0.7330 \pm 0.4353$ | $0.0313 \pm 0.0196$ | $0.1424 \pm 0.4681$ |
| $-10^\circ$ | $37.2295 \pm 5.1967$ | $0.4247 \pm 0.2557$ | $0.0283 \pm 0.0151$ | $0.0923 \pm 0.1618$ |
| $-5^\circ$ | $42.2295 \pm 5.1967$ | $0.3330 \pm 0.1876$ | $0.0276 \pm 0.0155$ | $0.0956 \pm 0.1849$ |
| $0^\circ$ | $47.2295 \pm 5.1967$ | $0.2906 \pm 0.1499$ | $0.0294 \pm 0.0165$ | $0.0000 \pm 0.0000$ |
| $+5^\circ$ | $52.2295 \pm 5.1967$ | $0.3100 \pm 0.1870$ | $0.0282 \pm 0.0152$ | $0.0732 \pm 0.1388$ |
| $+10^\circ$ | $57.2295 \pm 5.1967$ | $0.3923 \pm 0.2378$ | $0.0290 \pm 0.0159$ | $0.1164 \pm 0.3282$ |
| $+20^\circ$ | $67.2295 \pm 5.1967$ | $0.5905 \pm 0.3039$ | $0.0323 \pm 0.0169$ | $0.1233 \pm 0.2456$ |
| $+40^\circ$ | $87.2295 \pm 5.1967$ | $1.1158 \pm 0.4740$ | $0.0379 \pm 0.0239$ | $0.1554 \pm 0.3539$ |

After averaging positive and negative directions, centerline RMSE increased from 0.3215 mm at $|\delta| = 5^\circ$ to 0.4085, 0.6618, and 1.2832 mm at $|\delta| = 10^\circ$, $20^\circ$, and $40^\circ$; the corresponding absolute FFR deviations were 0.0844, 0.1044, 0.1329, and 0.1594. Radius error was comparatively stable within $\pm 10^\circ$, whereas centerline error increased earlier because angular departure directly weakens depth triangulation. Negative perturbations were more detrimental because they reduced the inter-view baseline; at $-40°$, 12 cases crossed the view-coincidence point.

Based on these results, each view should remain within 5° beyond the training distribution. In the source-side convention, the empirical operating ranges are RAO $40° \pm 15°$ with CRA/CAU $0° \pm 10°$ for the first view, and RAO/LAO $0° \pm 10°$ with CAU $25° \pm 15°$ for the second view.

### F.2 Segmentation perturbation

Segmentation uncertainty was propagated through the complete reconstruction and JacobiNet–PINN workflow in all 32 patients. Localized uncertainty was represented by mixed one-pixel dilation or erosion of randomly selected 10%, 50%, or 100% of boundary points. Deterministic one-pixel dilation or erosion of each entire view separately and of both views simultaneously served as progressively stronger systematic-bias stress tests. All other inputs and model settings were fixed.

**Table F2. End-to-end sensitivity to one-pixel segmentation perturbations.**

| Perturbation | Radius-profile MAE (mm) | Absolute *Δ*MLD (mm) | Absolute *Δ*FFR | Maximum absolute *Δ*FFR | FFR accuracy |
|---|---|---|---|---|---|
| Baseline | — | — | — | — | 93.8% |
| 10% boundary points, mixed dilation/erosion | $0.080 \pm 0.039$ | $0.154 \pm 0.104$ | $0.047 \pm 0.047$ | 0.173 | 93.8% |
| 50% boundary points, mixed dilation/erosion | $0.103 \pm 0.054$ | $0.169 \pm 0.145$ | $0.047 \pm 0.050$ | 0.205 | 90.6% |
| 100% boundary points, mixed dilation/erosion | $0.109 \pm 0.057$ | $0.187 \pm 0.158$ | $0.049 \pm 0.060$ | 0.278 | 87.5% |
| View A, all-point dilation | $0.084 \pm 0.034$ | $0.192 \pm 0.110$ | $0.043 \pm 0.049$ | 0.156 | 84.4% |
| View A, all-point erosion | $0.062 \pm 0.028$ | $0.170 \pm 0.102$ | $0.058 \pm 0.072$ | 0.278 | 87.5% |
| View B, all-point dilation | $0.079 \pm 0.031$ | $0.163 \pm 0.095$ | $0.031 \pm 0.037$ | 0.126 | 87.5% |
| View B, all-point erosion | $0.059 \pm 0.040$ | $0.136 \pm 0.083$ | $0.050 \pm 0.068$ | 0.278 | 90.6% |
| Both views, all-point dilation | $0.145 \pm 0.046$ | $0.319 \pm 0.120$ | $0.052 \pm 0.056$ | 0.194 | 78.1% |
| Both views, all-point erosion | $0.127 \pm 0.054$ | $0.301 \pm 0.116$ | $0.084 \pm 0.091$ | 0.284 | 81.3% |

Values are patient-level means $\pm$ sample SD. Under localized mixed perturbations, mean absolute FFR change remained 0.047–0.049 and accuracy remained 87.5%–93.8%. Uniform same-direction displacement of every boundary point, particularly in both views, produced larger changes and defines a conservative systematic-bias sensitivity limit rather than routine annotation variability.

### F.3 Paired CTA comparison

Paired CTA-derived minimal lumen diameter (MLD) and lesion length were available in five patients. No paired IVUS or OCT lumen data were available.

**Table F3. Reconstruction versus paired CTA measurements.**

| Study ID | Reconstructed MLD (mm) | CTA MLD (mm) | Absolute MLD difference (mm) | Reconstructed lesion length (mm) | CTA lesion length (mm) | Absolute length difference (mm) |
|---|---|---|---|---|---|---|
| 00001 | 1.518 | 1.532 | 0.014 | 13.078 | 14.554 | 1.476 |
| 00004 | 1.537 | 1.601 | 0.064 | 10.659 | 11.856 | 1.197 |
| 00021 | 1.207 | 1.297 | 0.090 | 11.375 | 10.975 | 0.400 |
| 00025 | 2.094 | 2.033 | 0.061 | 14.232 | 13.153 | 1.079 |

| Study ID | Reconstructed MLD (mm) | CTA MLD (mm) | Absolute MLD difference (mm) | Reconstructed lesion length (mm) | CTA lesion length (mm) | Absolute length difference (mm) |
|---|---|---|---|---|---|---|
| 00031 | 1.526 | 1.465 | 0.061 | 12.136 | 13.226 | 1.090 |

The MLD bias (reconstruction minus CTA) was $-0.009 \pm 0.070$ mm, with an MAE of 0.058 mm and an RMSE of 0.063 mm. Reconstructed and CTA lesion lengths were $12.296 \pm 1.407$ and $12.753 \pm 1.378$ mm, respectively. The lesion-length bias was $-0.457 \pm 1.127$ mm, with an MAE of 1.048 mm and an RMSE of 1.107 mm.

### F.4 Inter-observer reproducibility

A second observer independently re-annotated the same two angiographic views for all 32 patients. Both observers followed the same prespecified workflow: de-identified paired angiograms, lesion-region lumen delineation in ITK-SNAP, mask continuity and completeness checks, and mask locking before reconstruction. The second observer could not view the first observer's contours. Both annotation sets were generated without access to invasive FFR or downstream reconstruction and flow results. Imaging calibration, reconstruction weights, physiological inputs, and JacobiNet–PINN settings were unchanged.

**Table F4. Inter-observer agreement.**

| Endpoint | Original annotation | Independent re-annotation | Difference | Mean absolute difference | ICC(2,1) (95% CI) |
|---|---|---|---|---|---|
| Inlet diameter (mm) | $3.184 \pm 0.421$ | $3.058 \pm 0.391$ | $-0.126 \pm 0.138$ | 0.155 | 0.901 (0.807–0.945) |
| MLD (mm) | $1.697 \pm 0.481$ | $1.660 \pm 0.503$ | $-0.037 \pm 0.139$ | 0.110 | 0.958 (0.904–0.980) |
| DS (percentage points) | $47.083 \pm 11.404$ | $46.232 \pm 12.709$ | $-0.851 \pm 4.819$ | 3.499 | 0.920 (0.833–0.962) |
| JacobiNet–PINN FFR | $0.848 \pm 0.098$ | $0.853 \pm 0.093$ | $+0.005 \pm 0.034$ | 0.025 | 0.938 (0.885–0.970) |

The Bland–Altman limits of agreement for FFR were $-0.061$ to $0.071$. Classification agreement at FFR $\leq 0.80$ was 30/32 (93.8%; exact 95% CI, 79.2%–99.2%), with Cohen's $\kappa = 0.861$ (bootstrap 95% CI, 0.634–1.000). Both annotation sets independently achieved 30/32 diagnostic accuracy against invasive FFR.

### F.5 Eccentric stenosis and non-circular cross-sections

The current representation jointly predicts a three-dimensional centerline and one scalar radius at each centerline position. It can therefore represent an axis-offset eccentric stenosis with locally circular cross-sections, but it cannot directly encode circumferentially varying D-shaped, irregular, or multilobed sections. These two properties were evaluated separately.

Three paired synthetic geometries were constructed. Within each pair, the concentric and eccentric lesions had identical inlet and outlet sections and matched longitudinal radius and cross-sectional-area profiles; only the local centerline displacement differed. Analytic ground-truth and reconstructed geometries were evaluated using the same JacobiNet–PINN configuration at 0.383 m/s.

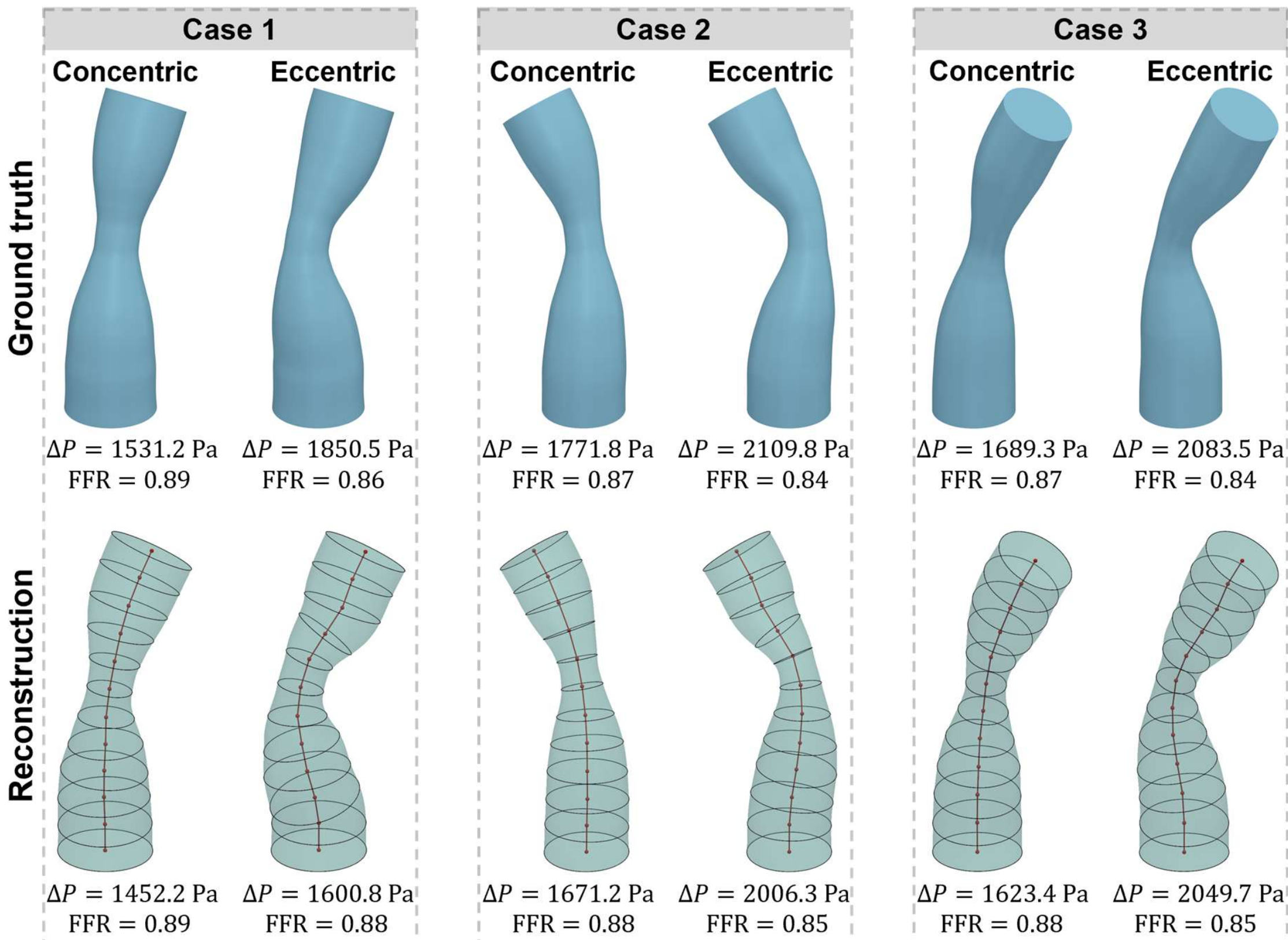

**Figure F1. Reconstruction and hemodynamic evaluation of concentric and eccentric stenoses.** The upper row shows paired ground-truth geometries, and the lower row shows their reconstructions.

**Table F5. Paired concentric and eccentric stenoses.**

| Case | Geometry | Analytic-GT $\Delta P$ (Pa) | Reconstructed $\Delta P$ (Pa) | Analytic-GT FFR | Reconstructed FFR |
|---|---|---|---|---|---|
| 1 | Concentric | 1,531.2 | 1,452.2 | 0.8852 | 0.8911 |
| 1 | Eccentric | 1,850.5 | 1,600.8 | 0.8612 | 0.8799 |
| 2 | Concentric | 1,771.8 | 1,671.2 | 0.8671 | 0.8747 |
| 2 | Eccentric | 2,109.8 | 2,006.3 | 0.8418 | 0.8495 |
| 3 | Concentric | 1,689.3 | 1,623.4 | 0.8733 | 0.8782 |
| 3 | Eccentric | 2,083.5 | 2,049.7 | 0.8437 | 0.8463 |

Eccentric displacement increased analytic pressure drop by 19.1%–23.3%. The reconstruction preserved the higher pressure drop and lower FFR in every eccentric pair. Across the six vessels, the maximum absolute DS error was 0.712 percentage points, pressure-drop MAPE was $5.792 \pm 4.038\%$, and absolute FFR error was $0.0079 \pm 0.0056$ (maximum, 0.0187).

The effect of non-circularity was isolated using area-matched CFD. The centerline and area at every axial position were fixed, while the circular throat was transformed into ellipses with minor-to-major axis ratios of 0.8, 0.6, and 0.4. Circular or near-circular and elliptical lumens represent the dominant coronary cross-sectional shapes reported by in vivo IVUS [44,45]. Figure F2 compares the CFD-derived pressure drop and FFR across the circular and area-matched elliptical throat geometries.

| Throat shape | Minor/major axis ratio | CFD ΔP (Pa) | CFD FFR |
|---|---|---|---|
| Circle | 1.00 | 1980.1 | 0.851 |
| Ellipse | 0.80 | 1984.3 | 0.851 |
| Ellipse | 0.60 | 2006.4 | 0.850 |
| Ellipse | 0.40 | 2069.9 | 0.845 |

**Figure F2. Sensitivity of pressure drop and FFR to area-matched throat shape.**

At a minor-to-major axis ratio of 0.4, pressure drop increased by 4.533% and FFR changed by 0.006732; changes at ratios of 0.8 and 0.6 were smaller. Global pressure drop and FFR were therefore relatively insensitive to the tested area-matched elliptical deformations. However, these findings should not be generalized to arbitrary irregular contours or circumferentially resolved local hemodynamics [46]; the associated limitations of the scalar-radius representation are discussed in the Discussion section of the main manuscript.